\documentclass[aps,prb,reprint,amsmath,amssymb,superscriptaddress,twocolumn,nofootinbib]{revtex4-2}
\usepackage{graphicx}
\usepackage{dcolumn}
\usepackage{bm}
\usepackage{physics}
\usepackage{graphicx}
\usepackage{subfigure}
\usepackage{braket}
\usepackage{amsmath}
\usepackage{commath}
\usepackage{mathtools}
\usepackage{chemformula}
\usepackage[normalem]{ulem}
\usepackage{wasysym}
\usepackage{txfonts}
\usepackage{tikz}

\begin{document}
	\preprint{APS/123-QED}
	\title{Current-Induced Circular Dichroism on Metallic Surfaces: A First-Principles Study}
        
	\author{Farzad Mahfouzi}
	\email{Farzad.Mahfouzi@nist.gov}
	\affiliation{Physical Measurement Laboratory, National Institute of Standards and Technology, Gaithersburg, MD 20899, USA.}
        \affiliation{Department of Chemistry and Biochemistry, University of Maryland, College Park, Maryland, MD 20742, USA} 
        \author{Mark D. Stiles}%
        \affiliation{Physical Measurement Laboratory, National Institute of Standards and Technology, Gaithersburg, MD 20899, USA.}	
	\author{Paul M. Haney}%
	\email{Paul.Haney@nist.gov}
        \affiliation{Physical Measurement Laboratory, National Institute of Standards and Technology, Gaithersburg, MD 20899, USA.}	
	\date{\today}
	
\begin{abstract}
We use {\it ab initio} calculations to understand the current-induced optical response and orbital moment accumulation at the surfaces of metallic films. These two quantities are related by a sum rule that equates the circular dichroic absorption integrated over frequency to the gauge-invariant self-rotation contribution to the orbital magnetization, $\vec{M}_{\rm SR}$.  In typical ferromagnets, $\vec{M}_{\rm SR}$ is a good approximation to the total orbital magnetization. We compute the current-induced $\vec{M}_{\rm SR}$ for a Pt thin film and compare it to the current-induced orbital moment accumulation calculated with the atom-centered approximation (ACA). We find significant differences: the size of $\vec{M}_{\rm SR}$ is, in general, larger than the ACA orbital moment accumulation by an order of magnitude and includes substantial finite-size effects. The differences between the two quantities caution against interpreting optical measurements with models utilizing the ACA. Finally, we compute the total $\vec{M}_{\rm SR}$ and ACA orbital moment accumulation as a function of layer thickness. For both quantities, the length scale at which the total surface accumulation saturates is on the order of the mean free path and longer than the length scale of their spatial profiles.

\end{abstract}
	
\maketitle
	
	
\section{\label{sec:sec1}Introduction}
	
Optical methods provide a direct and non-invasive way to measure a system's magnetic properties. Specifically, circular dichroism, which is the difference in a system's response to left-hand and right-hand circularly polarized light, indicates and can be used to characterize magnetic order. Circular dichroism includes optical phenomena such as the Magneto-optical Kerr effect (MOKE) and Faraday effect~\cite{OPPENEER2001_book}. The inherent simplicity of optical techniques makes them particularly valuable for probing subtle magnetic properties, including current-induced magnetization. Current-induced magnetization occurs in systems that lack inversion symmetry and is therefore present at sample surfaces in general~\cite{Yoda2018}.  Specific mechanisms for the current-induced magnetization at surfaces include the spin or orbital Edelstein effect~\cite{Yoda2018,Yoda2015,Go2017} and the spin or orbital moment accumulation due to a bulk spin or orbital Hall effect~\cite{DYAKONOV1971,  Hirsch1999, Sinova2015,  Bernevig2005, Kontani2008, Tanaka2008}.  Using MOKE measurements, the authors of Ref.~\cite{Kato2004} deduced the spin Hall conductivity and spin diffusion length of {\it n}-doped GaAs. Since then, numerous experiments have used similar methods and analyses to determine the spin Hall conductivity and spin diffusion length in a host of other materials~\cite{Stamm2017, Lyalin2023, Marui2024, Choi2023}.  These measurements use phenomenological models of spin\ diffusion to extract the spin accumulation and the corresponding characteristic length \cite{Stern2007}.

\begin{figure}
{\includegraphics[scale=1,angle=0,trim={0.0cm 0.0cm 0.0cm 1.0cm},clip,width=0.45\textwidth]{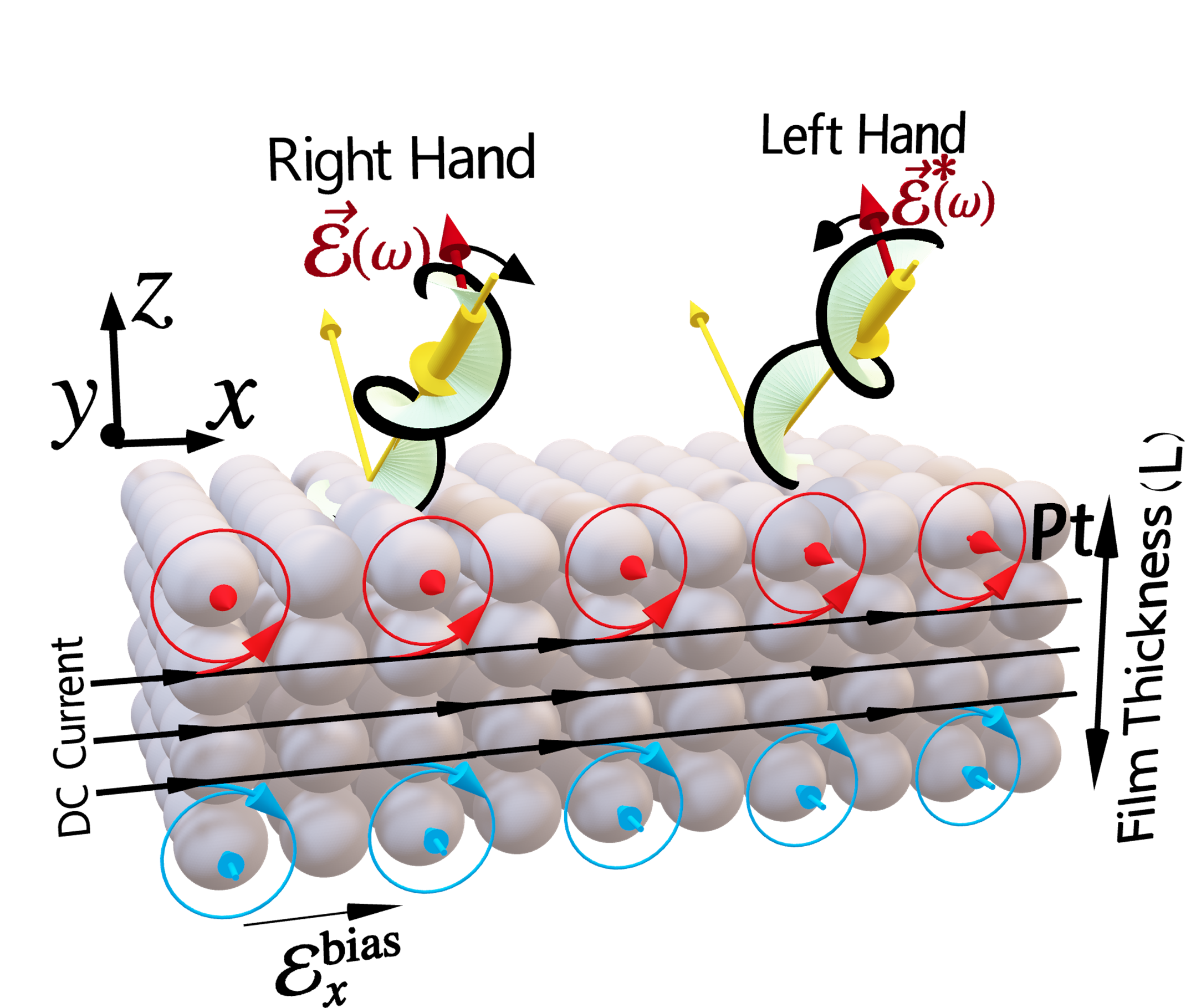}}
\caption{Schematic depiction of the current-induced circular dichroism on metallic surfaces. The bias direct current is along the $x$-axis.  In response, spin or orbital moment accumulation polarized perpendicular to the current flow direction and parallel to the surface planes, shown as blue and red circles are expected to occur, resulting in the breaking of time-reversal symmetry and circular dichroism. The yellow arrows depict the incident polarized light.   }
\label{fig:fig1}%
\end{figure}	

Recently, measured Kerr rotation signals have been used to study orbital moment transport and accumulation~\cite{Adamantopoulos2024, Park2012}. Doing so poses conceptual questions related to the unique and, at times, subtle properties of orbital magnetization.  For example, strong orbit-lattice coupling qualitatively modifies the nature of orbital moment transport~\cite{go2023long}.  Moreover, different descriptions of orbital magnetization are appropriate for different materials. In elemental transition metal ferromagnets, the orbital moment is well described by the atom-centered approximation (ACA)~\cite{Hanke2016}.  In this approximation, the orbital moment is taken to be confined near the atomic core.  However, for other systems, the ACA is not a good approximation for the orbital moment, and calculating the orbital moment requires the modern theory of orbital magnetization~\cite{Hanke2016}.  On a fundamental level, spatially-resolved descriptions of orbital magnetization density are {\it ill-defined} microscopically \cite{Bianco2013,seleznev2023towards}, but can be formulated on a sufficiently coarse length scale \cite{Bianco2013,jackson2012classical}. A formal description of the local surface orbital magnetization for the equilibrium ground state \cite{seleznev2023towards} has been recently developed, but it remains a challenge to develop a model that addresses the non-equilibrium current-induced orbital magnetization and can be incorporated into the analysis of optical experiments.  

Inspired by the questions raised by measurements of current-induced orbital moment accumulations with optical experiments~\cite{Stamm2017, Lyalin2023, Marui2024, Choi2023}, we compute the time-reversal odd part of the optoelectric response and compare it with the orbital moment accumulation of current-carrying Pt thin films. Calculating the optoelectric response avoids some of the difficulties associated with orbital magnetization described in the previous paragraph by directly computing what is measured. 

Optical measurements of the current-induced orbital moment accumulation at surfaces have been made with MOKE.  The comparison of these MOKE measurements with orbital moments is complicated by the dependence of the measured signal on material parameters besides the magnetization, such as the complex index of refraction. On the other hand, measurements of circular dichroism using X-ray magnetic circular dichroism (XMCD) \cite{thole1992x} have been used to determine the equilibrium orbital magnetization for transition metal ferromagnets \cite{chen1995experimental,kortright2000resonant} and other magnetic materials \cite{tietze2008xmcd,van1999orbital}. 

A first-principles calculation of current-induced MOKE is the subject of future work. We focus on calculating the circular dichroism in absorption rather than reflection since it is more directly related to orbital magnetization. The circular dichroic absorption integrated over all frequencies, $\vec{M}_{\rm SR}$, is formally identical to a portion of the ground-state orbital magnetization via a sum rule~\cite{oppeneer1998magneto,Souz2008}. Here, the SR subscript refers to the “self-rotation” part of the orbital magnetization, although this quantity only captures the gauge-invariant part of the self-rotation \cite{Souz2008}. For many materials studied to date, ${\vec M}_{\rm SR}$ is approximately equal to the total equilibrium orbital magnetic moment, although this need not be the case \cite{resta2020magnetic}.  While the applicability of this sum rule for non-equilibrium systems, like a current-induced moment, is not established, it nevertheless serves as an approximation to compare to other estimates of the non-equilibrium orbital magnetization.  

We compare our calculations of the orbital moment as captured by ${\vec M}_{\rm SR}$ to the current-induced orbital moment accumulation computed within the ACA. Much of the theoretical development of orbitronics relies on the ACA~\cite{go2020theory, Kontani2008, Bernevig2005}, making it essential to compare predictions derived within the ACA approximation with direct calculations of the measurable ${\vec M}_{\rm SR}$. Both calculations give contributions of opposite signs on the top and bottom surfaces of the thin film, indicating magnetic order, consistent with the picture of equal and opposite current-induced orbital moment accumulation on the surfaces (see Fig.~\ref{fig:fig1}). For thin film Pt, we find that $\vec{M}_{\rm SR}$ generally exceeds the ACA orbital moment accumulation by an order of magnitude. The spatial profile of both quantities is quite localized to the surface: their amplitudes decrease to less than $1/e$ of their maximum value within about $1~{\rm nm}$ from the surface. However, we find that as a function of film thickness, the total surface accumulation saturates over a different length scale. We find that this length is set by the mean free path and, for both quantities, is on the order of several nanometers.

The frequency-resolved ${\vec M}_{\rm SR}$ reveals two distinct contributions to the current-induced circular dichroic absorption. The first is from optical transitions between different bulk bands. The second is from transitions between subbands which result from the closed boundary condition along the film thickness direction. The inter-subband contribution to circular dichroic absorption dominates at lower frequencies and exhibits a longer characteristic length than the frequency-integrated quantity ${\vec M}_{\rm SR}$. The inter-subband response also exhibits a pronounced asymmetry in the conductivity tensor with respect to the electric field direction:  For a homogeneous electric field along the $z$-direction (out-of-plane), inter-subband transitions lead to equal and opposite in-plane currents at the top and bottom surfaces.  On the other hand, for a homogeneous electric field along the $x$-direction (in-plane), the inter-subband contribution to the response vanishes. The thickness dependence of the circular dichroic response at fixed frequency is quite different for responses dominated by bulk interband versus inter-subband transitions, which is described in more detail in this work.

The paper is organized as follows: Section~\ref{sec:sec2} presents the formalism describing the frequency-dependent conductivity, where we include the spatial dependence and nonlocal effects by computing the full 2-point response function.  The effects of a current bias are included by modifying the electron distribution function according to the relaxation time approximation. From the conductivity, we extract the circular dichroic absorption as a function of position.  In Section~\ref{sec:sec3}, we present results for the spatially-dependent longitudinal conductivity of a platinum film, the current-induced circular dichroism (Sec.~\ref{sec:sec3b}), and compare the self-rotation orbital moment accumulation to the accumulation computed with the ACA (sec.~\ref{sec:sec3c}). Appendix~\ref{sec.AppA} gives details of the computational methodology. Appendix~\ref{sec.AppB} explains the scaling of $\vec{M}_{\rm SR}$ at low frequencies as a function of film thickness. Appendix~\ref{sec.AppC} gives details of numerically extracting the thickness dependence of the optical response. Finally, Appendix~\ref{sec.AppD} gives the formal connection between the dichroic absorption and the circular photogalvanic effect.

\section{\label{sec:sec2}Theoretical Formalism}
		
In this section, we first derive the two-point frequency-dependent conductivity in equilibrium. The result is given in Eq.~\eqref{Eq.eq_opt_cond}, which is the standard expression for conductivity~\cite{Kubo1957,Greenwood1958} with the addition of site-projection operators on the perturbation and response matrix elements. Next, we review the sum rule which relates the frequency-integrated circular dichroic absorption to the gauge-invariant self-rotation part of the orbital magnetization $\vec{M}_{\rm SR}$, which is given by Eq.~\eqref{Eq.sum_rule_L2}. Finally, we generalize to non-equilibrium systems by including the change in occupation function due to the applied electric field, with the final result given in Eq.~\eqref{eq:alpha_eq}.

The total Hamiltonian of a periodic electronic system interacting with light is given by,
\begin{eqnarray}
{\mathbf{H}}_{\rm tot}(t)=\sum_{IJ{\mu\nu}}{\mathbf{c}}^{\dagger}_{I{\mu}}(t){\mathbf{H}}_{I{\mu},J{\nu}}(t){\mathbf{c}}_{J{\nu}}(t)
+\sum_{\alpha\vec{q}}\hbar\omega_{\vec{q}}\mathbf{a}^{\dagger}_{\alpha\vec{q}}(t)\mathbf{a}_{\alpha\vec{q}}(t),
\end{eqnarray}
where ${\mathbf{c}}^{\dagger}_{I{\mu}}(t)$ and ${\mathbf{c}}_{I{\mu}}(t)$ are the creation and annihilation operators for an electron at time $t$, atom $I$ and atomic orbital (including spin), ${\mu}$. Here, bold symbols represent operators in Fock (many-particle) space. The photonic dispersion is denoted by $\hbar\omega_{\vec{q}}$ which has dimensions of energy, and $\mathbf{a}^{\dagger}_{\alpha\vec{q}}(t)$ and $\mathbf{a}_{\alpha\vec{q}}(t)$ are the creation and annihilation operators for a photon with polarization $\alpha$, and momentum $\vec{q}$ at time $t$. The electronic tight-binding Hamiltonian between atoms $I$ and $J$ and orbitals $\mu$ and $\nu$ in a non-orthonormal atomic orbital basis set with overlap, ${\mathcal{O}}_{I{\mu},J{\nu}}$, is given by
\begin{align}
	{\mathbf{H}}_{I{\mu},J{\nu}}(t)=&\mathbf{H}_{I{\mu},J{\nu}}^0+\frac{e}{2}\left[\vec{\mathbf{A}}_{I}(t)\cdot\vec{{v}}_{I{\mu},J{\nu}}+\vec{{v}}_{I{\mu},J{\nu}}\cdot\vec{\mathbf{A}}_J(t)\right]\nonumber\\
 &+\frac{e^2}{4m}\left[{\mathbf{A}}_{I}^2(t)+{\mathbf{A}}_{J}^2(t)\right]{\mathcal{O}}_{I{\mu},J{\nu}},
 \end{align}
where $\mathbf{H}^0_{I{\mu}, J{\nu}}(t)$ is the electronic Hamiltonian in the absence of interactions with the photons, $\vec{{v}}_{I{\mu},J{\nu}}$ is the electronic velocity operator given in Fourier space by Eq.~\eqref{Eq.A1} in Appendix~\ref{sec.AppA}, $-e$ is the electron charge and $m$ is the electron mass. The electromagnetic vector potential at site $I$ is
\begin{align}
	\vec{\mathbf{A}}_I(t)=&\sqrt{\frac{\hbar}{2V\omega_{\vec{q}}\epsilon_0}}\sum_{\alpha\vec{q}}\left(\mathbf{a}^{\dagger}_{\alpha\vec{q}}(t)\vec{\rm e}^{\mbox{*}}_{\alpha\vec{q}}e^{i\vec{q}\cdot\vec{r}_I}+\mathbf{a}_{\alpha\vec{q}}(t)\vec{\rm e}_{\alpha\vec{q}}e^{-i\vec{q}\cdot\vec{r}_I}\right),
\end{align}
		
where $\epsilon_0$ is the vacuum permittivity, $V$ is the volume, and $\vec{\rm e}_{\alpha\vec{q}}$ is the photon polarization vector, which can be complex in general.  In the following, for compactness, we suppress the atomic and orbital indices, $I,{\mu}$, and represent the corresponding matrices by symbols with a hat. 
		
The Heisenberg equation of motion for the photonic creation/annihilation operators is
\begin{flalign}
i\frac{d\mathbf{a}_{\alpha\vec{q}}(t)}{dt}&=\frac{1}{\hbar}[\hat{\mathbf{H}}_{{\rm tot}}(t),\mathbf{a}_{\alpha\vec{q}}(t)]=\omega_{\vec{q}}\mathbf{a}_{\alpha\vec{q}}(t)\label{Eq.Eq4} \\
&+\frac{1}{2}\sqrt{\frac{ e^2}{2\hbar V\omega_{\vec{q}}\epsilon_0}}{\rm Tr}\left[\vec{\rm e}^{\mbox{*}}_{\alpha\vec{q}}\cdot\{\vec{\hat{v}},e^{i\vec{q}\cdot\vec{\hat{r}}}\}\hat{\boldsymbol{\rho}}(t)\right]\nonumber,
\end{flalign}
where $\vec{\hat{r}}$ is the position operator matrix, $\{,\}$ denotes the anticommutation and we ignored the contribution from the ``plasmon" ($\frac{e^2}{2m}\vec{\mathbf{A}}^2$) term since in this work we are interested in dichroic response and this term is insensitive to light polarization.  Given the density matrix at an initial time $t_0$, at any later time it follows the relation \mbox{$\hat{\boldsymbol{\rho}}(t)=\hat{\boldsymbol{G}}^r(t,t_0)\hat{\boldsymbol{\rho}}(t_0)\hat{\boldsymbol{G}}^a(t_0,t)$}, where $\hat{\boldsymbol{G}}^{r(a)}(t,t_0)$ is the total retarded (advanced) electronic Green's function. The total density matrix is given by \mbox{$\hat{\boldsymbol{\rho}}(t)=\delta\hat{\boldsymbol{\rho}}(t)+\int dE~ \hat{{\rho}}_0(E)$}, where the second term is the time-independent and does not contribute to Eq.~\eqref{Eq.Eq4}. 

A perturbative expansion of $\hat{\boldsymbol{G}}^{r(a)}$ to the lowest order in bosonic operators $\mathbf{a}_{\beta\vec{q}}$ results in the following expression
\begin{widetext}
\begin{flalign}
\hat{\boldsymbol{G}}^{r(a)}(E,E\pm\omega)=\frac{1}{2}\sqrt{\frac{\hbar e^2 }{2V\omega_{\vec{q}}\epsilon_0}}\sum_{\beta}\mathbf{a}_{\beta\vec{q}}(\omega) 
\hat{G}^{r(a)}_0(E)\{\vec{\rm e}_{\beta\vec{q}}\cdot\vec{\hat{v}},e^{-i\vec{q}\cdot\vec{\hat{r}}}\}\hat{G}^{r(a)}_0(E\pm\hbar\omega), \label{Eq.G_perturb}
\end{flalign}
where $\hat{G}_0^{r(a)}(E)$ is the Fourier transform of the unperturbed electronic Green's function. Note that the unperturbed electronic Green's function $\hat{G}_0^{r(a)}(t,t_0)$ depends only the time difference $t-t_0$, and can be Fourier transformed in terms of a single energy $E$: $\hat{G}_0^{r(a)}(E) = \int dt ~e^{iE(t-t_0)} \hat{G}_0^{r(a)}(t-t_0)$.  In contrast, the full Green's function $\hat{G}^{r(a)}(t,t_0)$ is time-dependent, and its Fourier transform requires two parameters: $\hat{G}^{r(a)}(E,E') = \int dt' \int dt ~e^{iE't_0}e^{iEt} ~\hat{G}^{r(a)}(t,t_0)$. Eq.~\eqref{Eq.G_perturb} yields the following nonequilibrium contribution to the density matrix
\begin{flalign}\label{eq.noneq_DM}
\delta\hat{\boldsymbol{\rho}}(\omega)=\int dt~e^{i\omega t}\delta\hat{\boldsymbol{\rho}}(t)=\frac{1}{2}
\sqrt{\frac{\hbar e^2 }{2V\omega_{\vec{q}\epsilon_0}}}\sum_{\beta}\mathbf{a}_{\beta\vec{q}}(\omega)\int dE~ \vec{\rm e}_{\beta\vec{q}}\cdot\left(\hat{G}^r_0(E)\{\vec{\hat{v}},e^{-i\vec{q}\cdot\vec{\hat{r}}}\}\hat{\rho}_0(E+\hbar\omega)+\hat{\rho}_0(E)\{\vec{\hat{v}},e^{-i\vec{q}\cdot\vec{\hat{r}}}\}\hat{G}^a_0(E+\hbar\omega)\right),
\end{flalign}
where $\hat{\rho}_0(E)={\rm Im}(\hat{G}_0^r(E))f(E)$ is the equilibrium density matrix. To obtain the photon propagation equation in the material, we take the Fourier transform of Eq.~\eqref{Eq.Eq4} and plug in the above expression for $\delta\hat{\boldsymbol{\rho}}(\omega)$ to obtain:
\begin{flalign}
(\omega-\omega_{\vec{q}})\mathbf{a}_{\alpha\vec{q}}(\omega)-i\frac{1}{\epsilon_0}\sum_{\beta}\vec{\rm e}^{\mbox{*}}_{\alpha\vec{q}}\cdot\hat{\sigma}(\omega,\vec{q})\cdot\vec{\rm e}_{\beta\vec{q}}\mathbf{a}_{\beta\vec{q}}(\omega)=0,
\end{flalign}
where $\hat{\sigma}$ is a 3$\times$3 conductivity tensor with elements that are given by

\begin{flalign}
\sigma_{j\ell}(\omega,\vec{q})={\frac{ e^2}{8i\pi V\omega}}\int dE~{\rm Tr}\left[\{{\hat{v}}^{j},e^{i\vec{q}\cdot\vec{\hat{r}}}\}\right.
\left.\left(\hat{G}_0^r(E)\{{\hat{v}}^{\ell},e^{-i\vec{q}\cdot\vec{\hat{r}}}\}\hat{\rho}_0(E+\hbar\omega)+\hat{\rho}_0(E)\{{\hat{v}}^{\ell},e^{-i\vec{q}\cdot\vec{\hat{r}}}\}\hat{G}^a_0(E+\hbar\omega\right)\right]\nonumber.
\end{flalign}
\end{widetext}
			
{\it Layer-Resolved Optical Conductivity.} 	
Using the ground-state electronic density matrix, the two-point linear response optical conductivity can be calculated using
\begin{flalign}\label{Eq.eq_opt_cond}
\sigma^{IJ}_{j\ell}(\omega)={\frac{ \hbar e^2}{2iVN_k}}\sum_{mn\vec{k}}&\frac{{{v}}^{j I}_{mn\vec{k}}{{v}}^{\ell J}_{nm\vec{k}}}{\varepsilon_{m\vec{k}}-\varepsilon_{n\vec{k}}-\hbar\omega-i\eta}\frac{f_{n\vec{k}}-f_{m\vec{k}}}{\hbar\omega}, 
\end{flalign}
where $N_k$ is the number of $k$-points in the Brillouin zone and we used the Bloch states basis set to diagonalize the Hamiltonian and carry out the integral over energy analytically. The Fermi distribution function is denoted $f_{n\vec{k}}=f(\varepsilon_{n\vec{k}})$, $V$ is the volume per atom and $\eta=$~50~meV is a constant energy broadening.  The layer-resolved velocity operator is defined as, $\vec{\hat{v}}^I=(\vec{\hat{v}}\hat{P}_I+\hat{P}_I\vec{\hat{v}})/2$, where $\hat{P}_I$ is the layer projection operator (i.e., an identity matrix for orbitals corresponding to layer $I$ and zero elsewhere). 
Here and in what follows, we neglect the in-plane spatial dependence and take the photon momentum $\vec{q}$ to be zero.

The two-point optical conductivity tensor can be understood as the nonlocal linear response function	
\begin{flalign}\label{Eq.Ohm}
{\rm\mathcal{J}}_{j I}(\omega)&=\sum_{J\ell}\sigma^{IJ}_{j\ell }(\omega)\mathrm{\rm \mathcal{E}}_{\ell J}(\omega) ,
\end{flalign}
which describes the current density, ${\rm \mathcal{J}}_{j I}$ in layer $I$ in response to an electric field ${\rm \mathcal{E}}_{\ell J}$ in layer $J$

The current-induced change of the optical conductivity can be obtained by including the electric field-induced change in the electron distribution function $\delta f$. Under the relaxation time approximation and ignoring the interband contribution, which has a negligible contribution when $\eta$ is much smaller than band energy splittings, this effect is expressed as \mbox{$\delta f_{n\vec{k}}=e\mathcal{E}^{\rm bias}_{\gamma}(\partial f_{n\vec{k}}/\partial k_{\gamma})/\eta$}, yielding,
\begin{flalign}
\chi^{IJ}_{j\ell;\gamma}(\omega)=&{\frac{\hbar e^2}{2iVN_k}}\sum_{mn\vec{k}}\frac{{{v}}^{j I}_{mn\vec{k}}{{v}}^{\ell J}_{nm\vec{k}}}{\varepsilon_{n\vec{k}}-\varepsilon_{m\vec{k}}+\hbar\omega-i\eta}\frac{\delta f^{\gamma}_{n\vec{k}}-\delta f^{\gamma}_{m\vec{k}}}{\hbar\omega},\label{Eq.neq_opt_cond_a}
\end{flalign}
where, \mbox{$\chi^{IJ}_{j\ell;\gamma}=\delta \sigma^{IJ}_{j\ell}/e\mathcal{E}^{\rm bias}_{\gamma}$} and, 
\begin{flalign}
\delta f^{\gamma}_{n\vec{k}}=\delta f_{n\vec{k}}/e\mathcal{E}^{\rm bias}_{\gamma}=\frac{1}{\eta}{v}^{\gamma}_{nn\vec{k}}f'(\varepsilon_{n\vec{k}}).
\end{flalign} 
Here, $f'$ is the derivative of the Fermi distribution function.
   
{The electronic power loss, equivalent to the rate of photonic energy absorption, is given by the real part of $A(\omega)=\sum_I\vec{\rm \mathcal{J}}_I(\omega)\cdot \vec{{\rm \mathcal{E}}}_I^{\mbox{*}}(\omega)$.  The absorption can be decomposed into four terms, as follows,
\begin{flalign}
A(\omega)&=\frac{1}{4}\sum_{IJj\ell}\left({\rm Re}({\rm \mathcal{E}}_{j I}^{\mbox{*}}{\rm \mathcal{E}}_{\ell J}+{\rm \mathcal{E}}_{j J}^{\mbox{*}}{\rm \mathcal{E}}_{\ell I}){\rm Re}(\sigma^{IJ}_{j\ell}+\sigma^{JI}_{j\ell})\right.\nonumber\\
&+{\rm Re}({\rm \mathcal{E}}_{j I}^{\mbox{*}}{\rm \mathcal{E}}_{\ell J}-{\rm \mathcal{E}}_{j J}^{\mbox{*}}{\rm \mathcal{E}}_{\ell I}){\rm Re}(\sigma^{IJ}_{j\ell}-\sigma^{JI}_{j\ell})\nonumber\\
&-{\rm Im}({\rm \mathcal{E}}_{j I}^{\mbox{*}}{\rm \mathcal{E}}_{\ell J}+{\rm \mathcal{E}}_{j J}^{\mbox{*}}{\rm \mathcal{E}}_{\ell I}){\rm Im}(\sigma^{IJ}_{j\ell}+\sigma^{JI}_{j\ell})\nonumber\\
&\left.-{\rm Im}({\rm \mathcal{E}}_{j I}^{\mbox{*}}{\rm \mathcal{E}}_{\ell J}-{\rm \mathcal{E}}_{j J}^{\mbox{*}}{\rm \mathcal{E}}_{\ell I}){\rm Im}(\sigma^{IJ}_{j\ell}-\sigma^{JI}_{j\ell})\right)
\label{Eq.Eq9}.
\end{flalign}
Here, the first term describes a mechanism of absorbing linearly polarized light that is symmetric with respect to the layer indices, $I$ and $J$, so the lowest order contribution is independent of the photonic wave-vector $\vec{q}$, while the second term is anti-symmetric in layer indices so the lowest order contribution is linear in $\vec{q}$. Similarly, the third and fourth terms correspond to the symmetric and anti-symmetric components of circularly polarized light absorption, respectively. In bulk materials, since the anti-symmetric components of absorption are proportional to the photonic wavevector they may be ignored in the optical frequency range, particularly within or below the visible spectrum.  
  
{\it Orbital magnetization and $f$-sum rule}.  The orbital magnetization can be expressed in multiple equivalent forms. Within the modern theory of orbital magnetization, the total moment is given by

\begin{flalign}
\vec{M}^{\rm tot}_{\rm orb}&=\frac{e}{2\hbar N_k}\sum_{\substack{n\vec{k} \\ m\neq n}}\left(\varepsilon_{n\vec{k}}+\varepsilon_{m\vec{k}}-2{E}_{\rm F}\right)\frac{{\rm Im}(\vec{{v}}_{mn\vec{k}}\times\vec{{v}}_{nm\vec{k}})}{\left(\varepsilon_{n\vec{k}}-\varepsilon_{m\vec{k}}\right)^2}f_{n\vec{k}},
\end{flalign}
where $E_{\rm F}$ is the electronic Fermi energy. 
Ref.~\cite{Souz2008} breaks the total orbital magnetization into a gauge-invariant segment of self-rotation term denoted $\vec{M}_{\rm SR}$ and a residual term.  The ground state $\vec{M}_{\rm SR}$ is equal to the frequency-integrated circular dichroic absorption via a sum rule~\cite{Souz2008}.  Below, we develop a position-dependent formulation of $\vec{M}_{\rm SR}$, although, as alluded to in the introduction, there are subtleties in position-dependent descriptions of orbital moments.  For this work, the physical significance we ascribe to position dependence is in terms of the nonlocal conductivity. 

For a circularly polarized light in $xz$-plane, using the electric field polarization direction $\vec{\rm \mathcal{E}}_I=|\vec{\rm \mathcal{E}}_I|(\vec{\rm {e}}_x+ i\vec{\rm {e}}_z)/\sqrt{2}$, the corresponding helicity-dependent power absorption, given by the third term in Eq.~\eqref{Eq.Eq9}, integrated over frequency  can be calculated from

\begin{flalign}
M_{\rm SR}^{IJ;y}&=\frac{eV}{2\pi e^2}\hbar\int_0^{\infty} d\omega~{\rm Im}\left(\sigma^{IJ}_{xz} + \sigma^{JI}_{xz}-\sigma^{IJ}_{zx} - \sigma^{JI}_{zx}\right) \label{Eq.sum_rule_L}, \\
&=\frac{e}{4\pi N_k}\sum_{nm\vec{k}} \left({{x}}^{I}_{nm\vec{k}}{{v}}^{z J}_{mn\vec{k}}-{{z}}^{I}_{nm\vec{k}}{{v}}^{x J}_{mn\vec{k}}\right)\times\label{Eq.sum_rule_L2} \\
&~~~~~~~~~~~~~\left(f_{n\vec{k}}-f_{m\vec{k}}\right)~{\rm Im}\left(\log(\varepsilon_{m\vec{k}} - \varepsilon_{n\vec{k}}+i\eta)\right)+\left(I\leftrightarrow J\right)\nonumber.
\end{flalign}

The interband matrix elements of the position operator projected on layer $I$, $\vec{ r}^I_{mn\vec{k}}(n\neq m)$, are calculated using \mbox{$\vec{{r}}^{ I}_{mn\vec{k}}=i\hbar\vec{{v}}^{ I}_{mn\vec{k}}{\rm Re} \left(1/(\varepsilon_{m\vec{k}}-\varepsilon_{n\vec{k}}+i\eta) \right)$}~\cite{footnote}.  Since, ${\rm Im}\left(\log(x+i\eta)\right)/\pi$ can be recognized as a smeared Heaviside step function, the right side of Eq.~\eqref{Eq.sum_rule_L} can be identified with the expectation value of the orbital angular momentum $-e\langle\hat{P}\vec{\hat{r}}\times \hat{Q}\vec{\hat{v}}\rangle/2$, where $\hat{P}$ projects onto occupied states and $\hat{Q}$ projects on to unoccupied states.  

The extension of Eq.~\eqref{Eq.sum_rule_L} to account for current-induced changes to $\vec{M}_{\rm SR}$ can be achieved similarly to Eq.~\eqref{Eq.neq_opt_cond_a}, where the Fermi distribution function is replaced with the nonequilibrium occupation at the Fermi surface, consistent with the semiclassical picture. In this case, we define the current-induced self-rotating orbital magnetization $\delta \vec{M}_{\rm SR}$ per applied electric field as
\begin{flalign}
\mathcal{\alpha}^{IJ,j\ell}_{\rm SR}=&\frac{\delta M^{IJ;j}_{\rm SR}}{e\mathcal{E}^{\rm bias}_{\ell}}\label{Eq.eq12}.
\end{flalign}
The relevant component of the circular dichroic absorption for the system shown in Fig.~\ref{fig:fig1} is given by
\begin{eqnarray}
\mathcal{\alpha}^{IJ,yx}_{\rm SR}&=&\frac{eV}{2\pi e^2}\hbar\int_0^{\infty} d\omega~{\rm Im}\left(\chi^{IJ}_{xz;x} + \chi^{JI}_{xz;x}-\chi^{IJ}_{zx;x} - \chi^{JI}_{zx;x}\right), \nonumber\\
&\equiv&\hbar\int_0^{\infty} d\omega ~\mathcal{L}^{IJ}_{y;x}(\omega),\label{eq:alpha_eq}
\end{eqnarray}
where $\mathcal{L}_{y;x}(\omega)$ is defined as the frequency-resolved, current-induced circular dichroic absorption.

To compare the circular dichroic response with the current-induced orbital moment accumulation within the ACA, we use the following form of Kubo's formula to calculate the electric field-induced ACA orbital moment  accumulation on each layer of the film,
\begin{flalign}
\alpha^{I,j\ell}_{\rm ACA}=\frac{\delta{M}^{I,j}_{\rm ACA}}{e\mathcal{E}_{\ell}^{\rm bias}}=&{\frac{-e}{4 m_e N_k}}\sum_{n\vec{k}}\frac{1}{\eta}{{v}}^{\ell}_{nn\vec{k}}[{P}_{I}{{L}}^{j}]_{nn\vec{k}}f'(\varepsilon_{n\vec{k}})\label{Eq.CIO_A},
\end{flalign}
where $\vec{\hat{L}}$ is the angular momentum operator defined in the atomic orbitals basis set.  Within the constant broadening approximation, the ACA orbital moment accumulation of Eq.~\eqref{Eq.CIO_A} is related to the intrinsic orbital Hall effect (computed within the ACA) and to the transfer of angular momentum from orbit to spin and lattice \cite{go2020theory}.

To illustrate the use of the ACA and $\vec{M}_{\rm SR}$ as approximations for the total orbital moment, Fig.~\ref{fig:fig2}(a) shows ground state orbital moment of bulk Fe using both approximations and the full modern theory result. The results are consistent with prior {\it ab initio} investigations \cite{Hanke2016, Lopez2012} with some differences between our results and the earlier calculations. We find agreement between ACA and the total orbital moment near the Fermi energy but a relatively larger deviation when the chemical potential is distant from the Fermi level. We find that the ACA orbital moment and $\vec{M}_{\rm SR}$ approximations are similar to the full calculation close to the true Fermi energy. The agreement between these approaches demonstrates the reliability of the ACA when orbital magnetization arises from spin-orbit coupling to a collocated spin moment near the atomic core.

In the results of the next section, we analyze quantities derived from the two-point conductivity tensor.  Here, we introduce some key definitions. The local response at layer $I$ to a spatially uniform electric field is denoted by
\begin{eqnarray}
\sigma^{I,{\rm tot}}_{j\ell} &=& \sum_{J} \sigma^{IJ}_{j\ell}.\label{Eq.local_res}
\end{eqnarray}
To isolate the response at the top and bottom surface of the film, we consider the response of half the film to a spatially uniform electric field, which we denote by
\begin{eqnarray}\label{Eq.surf_res}
\sigma^{\rm surf}_{j\ell} &=& \sum_{I=1}^{N/2} \sigma^{I, {\rm tot}}_{j\ell}.
\end{eqnarray}
We make similar definitions for the current-induced change response function, $\chi_{j\ell;\gamma}^{I, {\rm tot}}$, $\chi_{j\ell;\gamma}^{\rm surf}$, $\alpha^{I,{\rm tot};j\ell}_{\rm SR}$ and  $\alpha^{{\rm surf},j\ell}_{\rm SR}$ { where to keep the Hermitian (dissipative) part of the current-induced optical conductivity tensor and consistent with Eq.~\eqref{Eq.sum_rule_L}, we anti-symmetrize the tensors with respect to the subscripts, $j$ and $\ell$}.\\

{\it Connection to other optoelectronic effects}.  From a broader perspective, current-induced circular dichroism is related to other optoelectronic effects, such as the circular photogalvanic effect.  In the circular photogalvanic effect, the absorption of circularly polarized light generates a direct charge current~\cite{Ganichev_2003}. { It occurs in materials or systems that lack inversion symmetry and has been used to study features of electronic structure such as Rashba spin-orbit splitting \cite{Olbrich2009, Karch_2010, Karch2011, Yuan2014, McIver2012, Okada2016,ivchenko2012superlattices}. {In centrosymmetric materials, this effect can be present at the metallic surfaces, resulting in the surface photo-galvanic effect,~\cite{perovich1981, Gurevich1993,magarill1981} which is attributed to the diffuse scattering of the photoexcited carriers in the subsurface layers~\cite{Mikheev2018}. } 

Both current-induced dichroism and the circular photogalvanic effect are second-order optical effects: in current-induced circular dichroism, a constant electric field mixes with an oscillating electric field to generate an oscillating response, while in the circular photogalvanic effect, two oscillating fields mix to generate a DC response. In this regard, the two effects are inversely related. Furthermore, as shown in Appendix~\ref{sec.AppD}, in the limit of small frequency ($\hbar\omega< kT$), the nonlinear response tensor elements of both phenomena obey identical expressions in terms of the electronic structure of the material, suggesting a shared microscopic origin between current-induced circular dichroism and the circular photogalvanic effect.}

\section{\label{sec:sec3}Results and Discussion}
\begin{figure}%
{\includegraphics[scale=0.32,angle=0,trim={0.5cm 0.5cm 0.0cm 0.0cm},clip,width=0.5\textwidth]{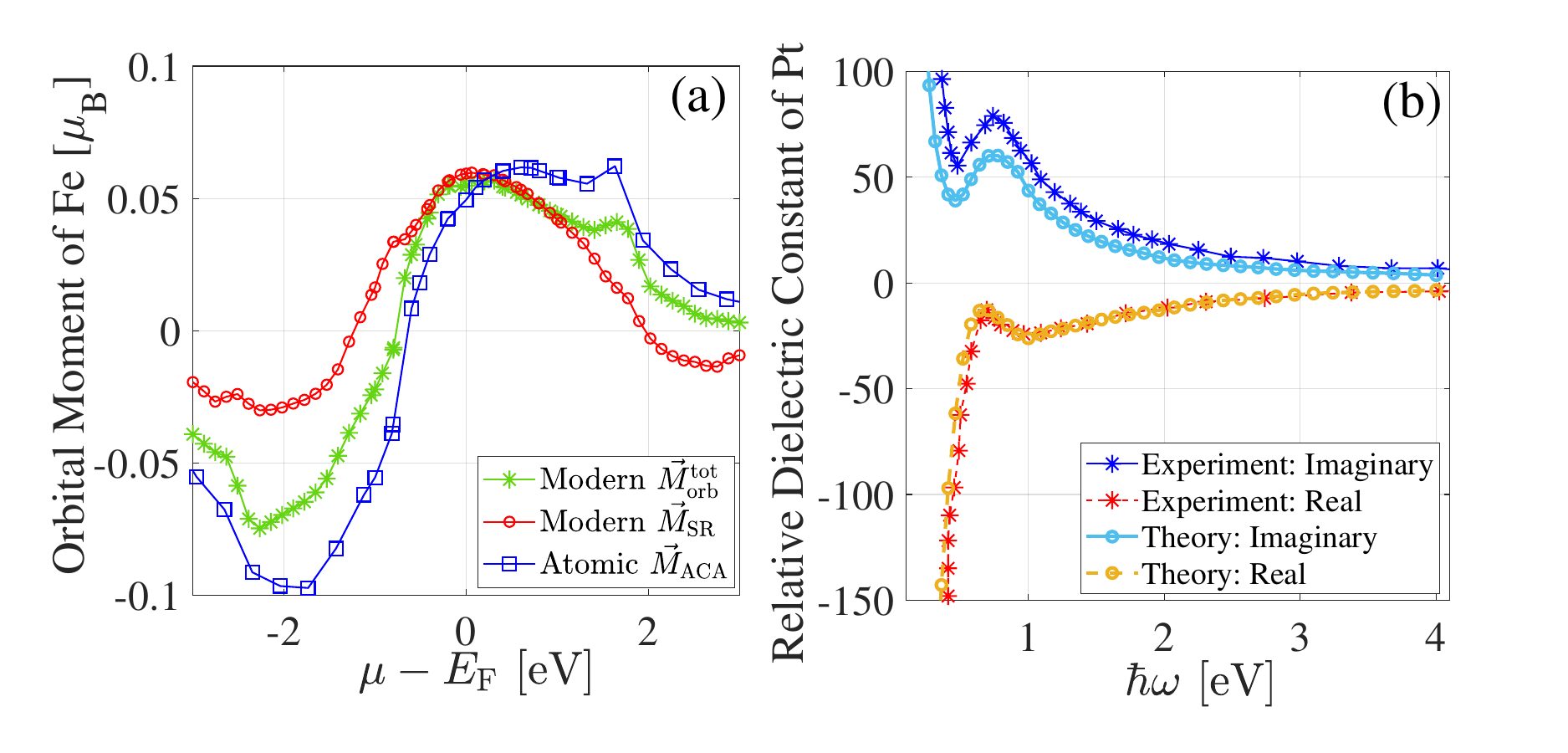}}%
\caption{(a) Comparison of the theoretically calculated total\mbox{(\textbf{\textcolor[rgb]{0.47,0.67,0.19}{---\raisebox{-0.25ex}{\scalebox{1.3}{$\mathclap{\mathclap{+}\mathclap{\times}}$}}---})}}, self-rotating\mbox{(\textbf{\textcolor[rgb]{1,0,0}{---\raisebox{-0.5ex}{\scalebox{1.75}{$\mathclap{\circ}$}}---})}}, and ACA\mbox{(\textbf{\textcolor[rgb]{0,0,1}{---\raisebox{-0.2ex}{\scalebox{1.2}{$\mathclap{\square}$}}---})}} components of the orbital moment in bulk Fe. (b) Comparison of the theoretical (lines with open circle symbols) and experimental (lines with star symbols) for the real (solid lines) and imaginary (dashed lines) parts of the optical conductivity for the central layer of a 37 monolayer Pt film. The experimental data are taken from Ref.~\cite{Weaver1975}.}%
\label{fig:fig2}%
\end{figure}
 As a case study, in what follows we consider Pt film for the numerical calculations. The geometry of the [001] Pt slab and the direction of the in-plane current bias (along the $x$-axis) are illustrated in Fig.~\ref{fig:fig1}. The direction normal to the plane is along the $z$ axis. We use a constant energy broadening $\eta=$~50~meV, corresponding to a relaxation time of $\tau=\hbar/(2\eta)=6.6~{\rm fs}$. This falls within the estimated range for bulk Pt, where $\tau$ ranges from $5.0~{\rm fs}$ to $8.0~{\rm fs}$ \cite{Dutta2017}.  Technical details of the {\it ab initio} calculation are found in Appendix~\ref{sec.AppA}.

  \subsection{\label{sec:sec3a}Linear Response Optical conductivity}

We begin by examining the position-dependent longitudinal conductivity of a Pt film in the absence of a bias current. For this calculation, we considered a film with 37 monolayers (ML). Calculating longitudinal conductivity is useful for illustrating the properties of the two-point conductivity response function and the influence of surfaces. In Fig.~\ref{fig:fig2}(b), we present both real and imaginary parts of the relative dielectric constant, \mbox{$\epsilon^r_{\alpha \beta}=\delta_{\alpha\beta}+i\sigma_{\alpha \beta}/\omega\epsilon_0$}, for the central layer of the Pt film as a function of frequency. We compare with experimental data~\cite{Weaver1975} and we find good agreement over the frequency range of the data. 

In Fig.~\ref{fig:fig3}(a), we present the real (dissipative) part of the calculated two-point optical conductivity for $\sigma_{xx}$ at fixed frequency $\hbar\omega=2~{\rm eV}$. We observe an approximately homogeneous and relatively local optical response. Fig.~\ref{fig:fig3}(b) shows the layer-resolved response to a uniform excitation. Near the surface, there is anisotropy between the in-plane and out-of-plane conductivity. Fig.~\ref{fig:fig3}(c) shows the position-dependent response for a localized perturbation at the middle layer. This represents the current induced in different layers $I$ for a fixed perturbation position at $J=(N+1)/2$.  The response is relatively local, extending a few layers away from the perturbation position.

\begin{figure} 
{\includegraphics[scale=0.32,angle=0,trim={0.8cm 0.2cm 1.0cm 0.0cm},clip,width=0.5\textwidth]{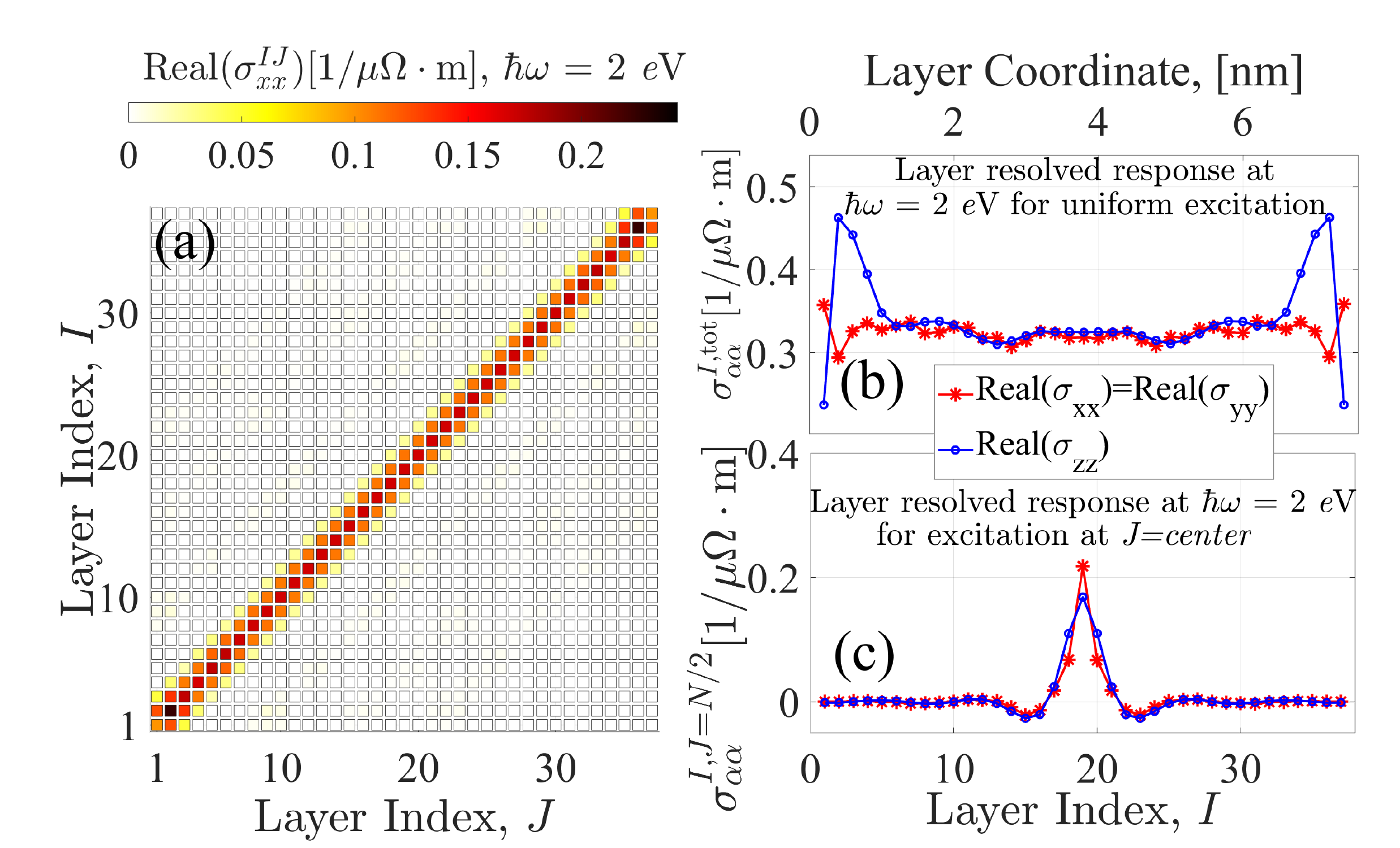}}
\caption{(a) Real part of the two-point optical conductivity tensor, $\sigma_{xx}^{IJ}$ for 37 monolayers Pt film. (b) The real part of the layer-resolved optical conductivity $\sum_J \sigma_{xx}^{IJ} = \sum_J \sigma_{yy}^{IJ}$ versus layer index, $I$.  (c) The nonlocal spread of the two-point optical conductivity tensor $\sigma_{xx}^{IJ}$ with $J=N/2$ for the central layer index versus the $I$-index. The red (blue) line with star (circle) symbols in (b) and (c) correspond to the in-plane (out-of-plane) component of the optical conductivity.}
\label{fig:fig3}
\end{figure}

\begin{figure} 
{\includegraphics[scale=0.32,angle=0,trim={0.5cm 0.8cm 0.0cm 0.0cm},clip,width=0.5\textwidth]{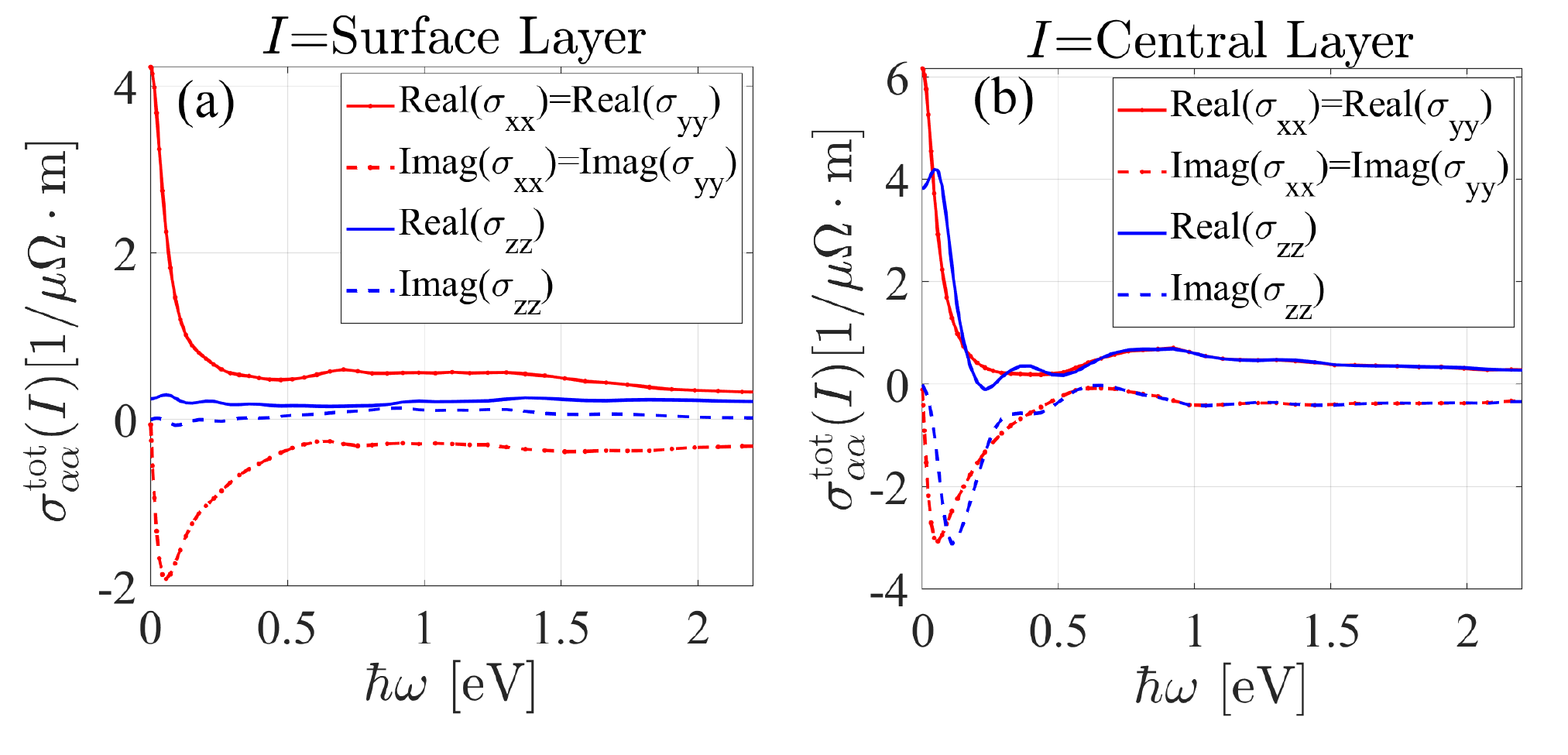}}
\caption{ Real and imaginary parts of the layer-resolved longitudinal optical conductivity tensor elements of a 37 monolayer Pt film versus frequency for (a) the surface layers and (b) the central layers.}
\label{fig:fig4}
\end{figure}

Next, we consider the frequency-dependent, layer-resolved conductivity for a uniform electric field. Fig.~\ref{fig:fig4}(a) and (b) show the results for surface and center layers, respectively. At the surface, the conductivity exhibits a large anisotropy between in-plane and out-of-plane response. The out-of-plane conductivity is reduced due to the suppression of out-of-plane current flow at the surface. In contrast, the in-plane conductivity exhibits a peak at low-frequency. This peak is due to transitions between inter-subband modes associated with the finite thickness of the film. The energy range of these modes, and therefore the width of this peak, is inversely proportional to $L$. In the limit of large $L$, this peak corresponds to the intraband Drude contribution to the conductivity.   

For the central layer response, the anisotropy between in-plane and out-of-plane conductivity is smaller than at the surface.  Again, we observe low-frequency peaks in both in-plane and out-of-plane conductivity components due to the intersubband transitions.  We remark that in the limit of $\omega\rightarrow 0$, the out-of-plane conductivity should vanish due to the closed boundary condition of the finite film.  However, we obtain a nonzero value due to finite smearing, {which is an artifact of the constant broadening approximation.}

\subsection{\label{sec:sec3b}Current-Induced Circular Dichroism}
\begin{figure} [tbp]
{\includegraphics[scale=0.32,angle=0,trim={0.0cm 1.0cm 1.0cm 0.0cm},clip,width=0.5\textwidth]{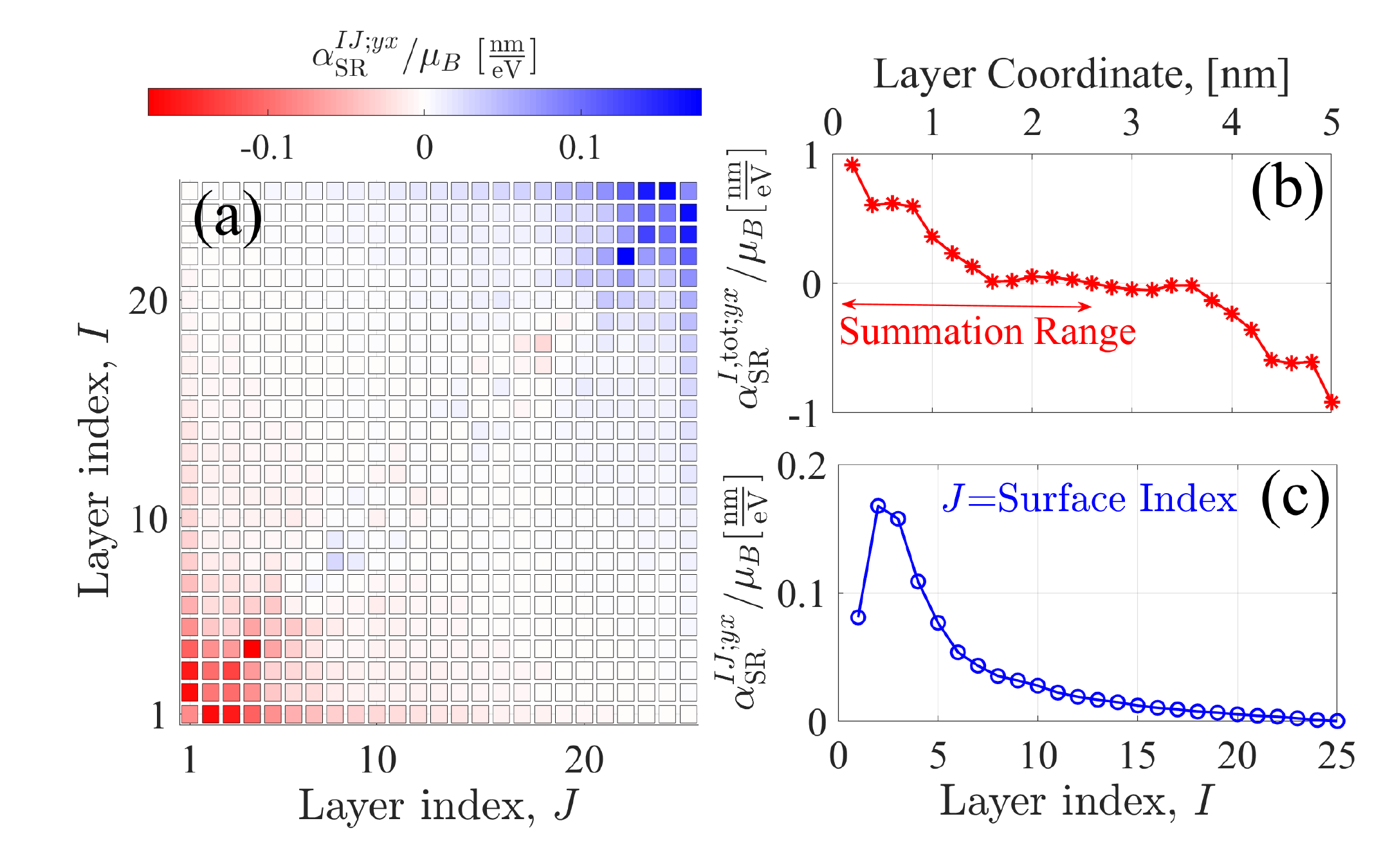}}
\caption{(a) Two-point, frequency-integrated, current-induced optical dichroic absorption $\alpha_{\rm SR}^{IJ;yx}$ for a 25 monolayer Pt film calculated using Eq.~\eqref{Eq.sum_rule_L}.  (b) Layer-resolved $\alpha_{\rm SR}^{I,{\rm tot};yx}$ versus layer index, calculated using ${\alpha}_{\rm SR}^{J,{\rm tot};\alpha\beta}=\sum_I{\alpha}_{\rm SR}^{IJ,\alpha\beta}$, following Eq.~\eqref{Eq.local_res}. (c) Nonlocal $\alpha_{\rm SR}^{IJ;yx}$ response between the $J$=surface layer and the other layers with index, $I$.}
\label{fig:fig5}
\end{figure}

We next consider the current-induced circular dichroic absorption in the Pt film.  Fig.~\ref{fig:fig5}(a) shows the two-point response function ${\alpha}_{\rm SR}^{IJ;yx} = \delta M^{IJ;y}_{\rm SR}/e\mathcal{E}_x^{\rm bias}$. As expected, the response is equal and opposite in the vicinity of the top and bottom surfaces. Compared to the longitudinal conductivity, the current-induced circular dichroism response is more nonlocal, where a response is induced several atomic layers away from a perturbation.  The position-dependent, current-induced $\vec{M}_{\rm SR}$ is shown in Fig.~\ref{fig:fig5}(b). The summation range that is used to calculate the optical response from half the film is also shown, which we refer to as the ``surface'' circular dichroic response given by Eq.~\eqref{Eq.surf_res}.  Fig.~\ref{fig:fig5}(c) shows a notably long decay length for the two-point circular dichroic response when one of the layer indices is fixed at the surface layer. This describes the presence of an in-plane AC current primarily on the surface layer in response to an out-of-plane AC electric field on a layer relatively far from the surface.

Figure~\ref{fig:fig6}(a) shows ${\alpha}^{{\rm surf},yx}_{\rm SR}$ for various film thicknesses as a function of an artificially shifted chemical potential, $\mu$, relative to the actual Fermi level $E_{\rm F}$.  We note that at sufficiently large thickness (compare 37 MLs with 55 MLs),  ${\alpha}^{{\rm surf},yx}_{\rm SR}$ saturates and becomes relatively thickness-independent.  Figure~\ref{fig:fig6}(b) shows the frequency-dependent circular dichroic absorption for these three thicknesses. At lower frequencies $\hbar\omega<1~{\rm eV}$, there is much stronger thickness dependence at fixed frequencies than might be expected from the frequency integrated behavior seen in Fig.~\ref{fig:fig6}(a). In contrast, at higher frequencies, the surface current-induced circular dichroism rapidly saturates as the film thickness exceeds 55 monolayers (approximately 11~nm), (see the inset of Fig.\ref{fig:fig6}(b)). At higher frequencies ($\hbar\omega\gtrapprox 2$ eV), a monotonically increasing behavior before the saturation with respect to the film thickness is not consistently observed. In Fig.~\ref{fig:fig6}(c), we plot $\rm{Im}(\chi_{zx;x}^{\rm surf})$, one of the terms comprising $\mathcal{L}_{y;x}^{\rm surf}$ (see Eq.~\eqref{eq:alpha_eq}). As discussed in Appendix~\ref{sec.AppB}, this term does not include contributions from inter-subband transitions. The value is considerably less than $\mathcal{L}_{y;x}^{\rm surf}$, indicating a large contribution from inter-subband transitions to $\mathcal{L}_{y;x}^{\rm surf}$ and strong asymmetry in the off-diagonal optical response.

\begin{figure}
{\includegraphics[scale=0.32,angle=0,trim={0.9cm 0.5cm 1.0cm 0.0cm},clip,width=0.5\textwidth]{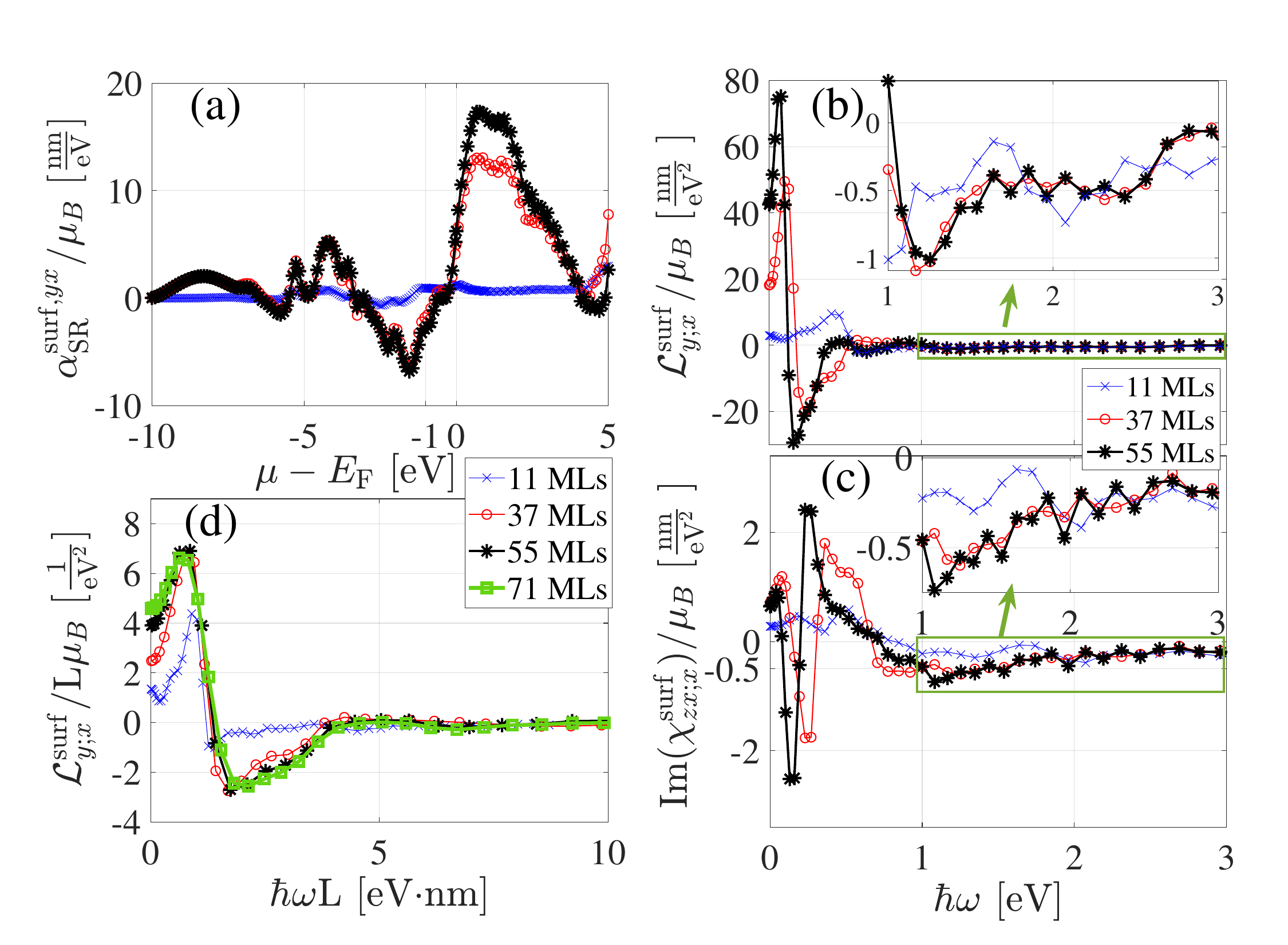}}%
\caption{(a) Current-induced $\vec{M}_{\rm SR}$ for half of the film versus a rigid shift of the chemical potential, $\mu$, for 11 \mbox{(\textbf{\textcolor[rgb]{0,0,1}{---\raisebox{-0.2ex}{\scalebox{1.2}{$\mathclap{\times}$}}---})}}, 37\mbox{(\textbf{\textcolor[rgb]{1,0,0}{---\raisebox{-0.5ex}{\scalebox{1.75}{$\mathclap{\circ}$}}---})}}, 55\mbox{(\textbf{\textcolor[rgb]{0,0,0}{---\raisebox{-0.25ex}{\scalebox{1.3}{$\mathclap{\mathclap{+}\mathclap{\times}}$}}---})}} and 71 \mbox{(\textbf{\textcolor[rgb]{0,1,0}{---\raisebox{-0.2ex}{\scalebox{1.2}{$\mathclap{\square}$}}---})}} monolayers of Pt film. (b) The imaginary part of the circular dichroic response, $\mathcal{L}^{IJ}_{y;x}={\rm Im}(\chi_{zx,x}^{IJ}-\chi_{xz,x}^{IJ})$, summed over half of the film versus frequency. (c) The imaginary part of the $zx$-component of the current induced optical response, which is more than an order of magnitude smaller than $\mathcal{L}^{IJ}_{y;x}$. (d) The rescaled plot of the circular dichroic response divided by the film thickness, $L$, versus $\hbar\omega L$. One monolayer is 0.198~nm.}
\label{fig:fig6}
\end{figure}

The behavior of $\mathcal{L}_{y;x}^{\rm surf}(\omega)$ at low frequencies ($\hbar\omega < 1~{\rm eV}$) is due to inter-subband transitions associated with the finite film thickness. The energies for which finite thickness effects occur are expected to scale as:
\begin{flalign}
    \hbar\omega <  \frac{\pi \hbar v_F}{L}\label{Eq.omegaL}
\end{flalign}
  where $v_F$ is the Fermi velocity.  Eq.~\eqref{Eq.omegaL} has a simple interpretation: $L/v_F$ is the time it takes for electrons to travel from one film surface to the other.  For perturbation periods longer than this transit time, electrons interact with both top and bottom surfaces.

Although the frequency-integrated dichroic response ${\alpha}^{{\rm surf},yx}_{\rm SR}$ becomes relatively independent of the film thickness for films thicker than approximately 55 monolayers (11~nm) (see Fig. \ref{fig:fig6}(a)), the spectral dichroic function $\mathcal{L}_{y;x}^{\rm surf}(\omega)$ remains sensitive to the thickness $L$, particularly in the low-frequency regime (see Fig. \ref{fig:fig6}(b)). Numerically, we find that for films thicker than the electronic mean-free path, $L\gg\ell_{\rm mfp}$, the full frequency and thickness dependence of $\mathcal{L}_{y;x}^{\rm surf}(\omega,L)$ is of the form 

  \begin{flalign}
  \mathcal{L}_{y;x}^{\rm surf}(\omega,L)
			\approx
   \mathcal{L}_{y;x}^{\rm surf,(0)}(\omega)+ L~\mathcal{L}_{y;x}^{\rm surf,(1)}(\omega L)\label{Eq.dich_opt_decomp},
		\end{flalign}	
Appendix~\ref{sec.AppB} discusses the origins of this form. The first term corresponds to a bulk response, while the second term is related to finite thickness effects. Fig.~\ref{fig:fig6}(d) shows $\mathcal{L}_{y;x}^{\rm surf}(\omega)$ divided by the film thickness $L$ versus $\hbar\omega L$. We observe significantly less sensitivity to the film thickness for $L>11$ nm,
indicating the dominance of the second term in Eq.~\eqref{Eq.dich_opt_decomp} for $\hbar\omega L < 4~{\rm eV\cdot nm}$. As discussed in Appendix~\ref{sec.AppB} and depicted in Fig.~\ref{fig:fig1}, the origin of the second term in Eq.~\eqref{Eq.dich_opt_decomp} lies in the inter-subband transitions.

\begin{figure}
{\includegraphics[scale=0.32,angle=0,trim={2.5cm 2.0cm 1.50cm 0.0cm},clip,width=0.5\textwidth]{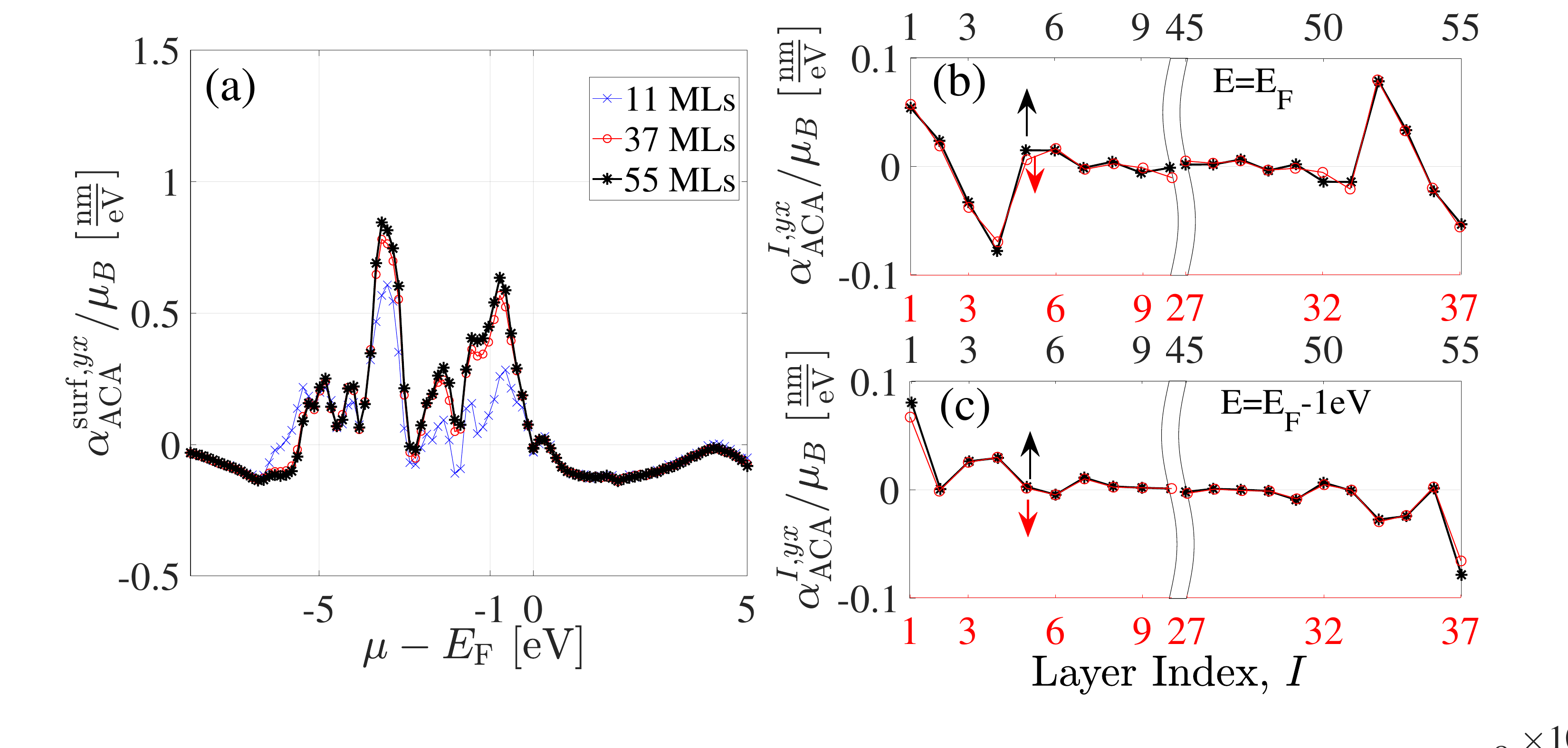}}
\caption{(a) Current-induced ACA orbital moment accumulation summed over half of the film versus a rigid chemical potential shift for 11 \mbox{(\textbf{\textcolor[rgb]{0,0,1}{---\raisebox{-0.2ex}{\scalebox{1.2}{$\mathclap{\times}$}}---})}}, 37\mbox{(\textbf{\textcolor[rgb]{1,0,0}{---\raisebox{-0.5ex}{\scalebox{1.75}{$\mathclap{\circ}$}}---})}} and 55\mbox{(\textbf{\textcolor[rgb]{0,0,0}{---\raisebox{-0.25ex}{\scalebox{1.3}{$\mathclap{\mathclap{+}\mathclap{\times}}$}}---})}} monolayers. (b) Layer resolved orbital moment accumulation for 37 and 55 monolayer Pt films at the Fermi level plotted versus layer index shown on the bottom and top horizontal axis, respectively. (c) Layer-resolved orbital moment accumulation for 37 and 55 MLs Pt film at the chemical potential shifted 1 eV below the Fermi level plotted versus the corresponding layer indices. }
\label{fig:fig7}
\end{figure}

  \subsection{\label{sec:sec3c}Comparison to current-induced ACA orbital moment accumulation}

We have considered the optical response associated with current-induced orbital moment accumulation at film surfaces.  It is of interest to correlate the optical response with a direct calculation of the total orbital moment accumulation.  As discussed in the introduction, computing the nonequilibrium, spatially-dependent orbital magnetization within the modern theory remains a complex problem.  Therefore, we opt for the commonly used ACA description of orbital magnetization.

In Fig.~\ref{fig:fig7}(a), we present the results for the ACA orbital moment accumulated on the film's bottom half versus the chemical potential. The findings reveal that at the Fermi level, the current-induced ACA orbital moment accumulation is nearly absent, and below the Fermi level, its magnitude varies with Pt thickness, growing as the film thickness increases and ultimately saturating. A direct comparison of Fig.~\ref{fig:fig7}(a) and Fig.~\ref{fig:fig6}(a) shows little correlation between $\vec{M}_{\rm SR}$ and the ACA orbital moment accumulation. On average, ${\alpha}^{{\rm surf},yx}_{\rm SR}$ exceeds ${\alpha}^{{\rm surf},yx}_{\rm ACA}$ by an order of magnitude. This is one of the primary findings of this work. We also observe another noticeable difference in the thickness dependence of ${\alpha}^{{\rm surf},yx}_{\rm SR}$ and ${\alpha}^{{\rm surf},yx}_{\rm ACA}$. At the smallest thickness of 11 ML, $\alpha_{\rm SR}^{{\rm surf},yx}$ is strongly suppressed at all chemical potentials (see Fig.~\ref{fig:fig6}(a)), whereas $\alpha_{\rm ACA}^{{\rm surf},yx}$ remains substantial for a wide range of chemical potentials (see Fig.~\ref{fig:fig7}(a)). This hints at a more localized, Edelstein-type behavior of the ACA orbital moment accumulation as compared to the dichroic optical response.

Fig.~\ref{fig:fig7}(b) presents the layer-resolved ACA orbital moment accumulation on the surface of the Pt films with 37 and 55 monolayer thickness. We observe that the current-induced ACA orbital moment accumulation is primarily localized near the surface layer of Pt and oscillates with the layer index, resulting in a near-zero total orbital moment accumulation. This finding is consistent with previous {\it ab initio} calculations \cite{Belashchenko2023,mahfouzi2020,rang2024}. In Fig.~\ref{fig:fig7}(c), we present the same results at a shifted chemical potential, 
$\mu=E_{\rm F}-1$ eV, where the total surface orbital moment accumulation is significant, allowing us to investigate the thickness dependence of this effect.

\begin{figure}
	{\includegraphics[scale=0.32,angle=0,trim={1.2cm 0.5cm 0.0cm 0.0cm},clip,width=0.5\textwidth]{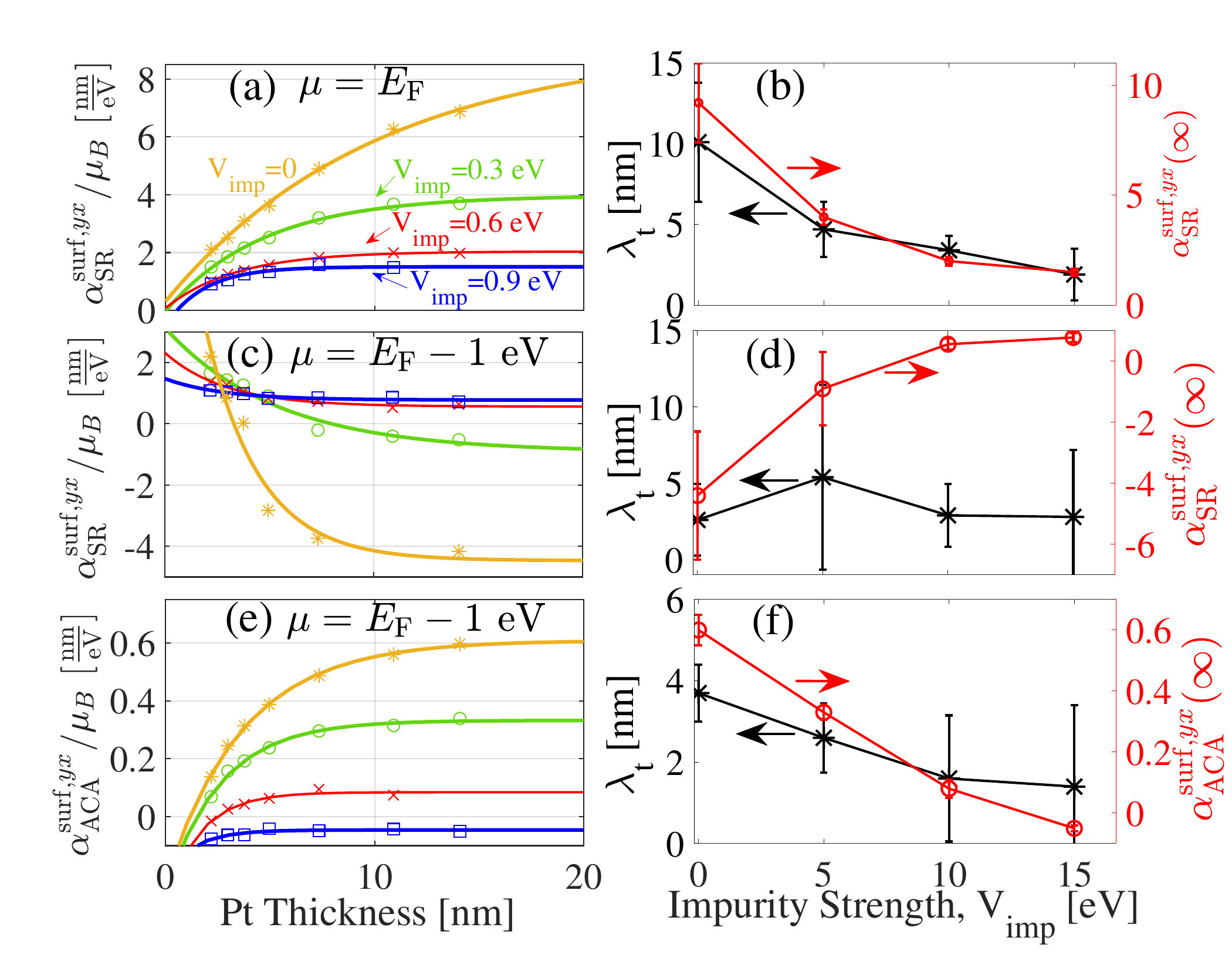}}
	\caption{ (a); Pt thickness dependence of the current-induced $M^y_{\rm SR}$ at the Fermi level, for various values of the impurity strengths, $V_{\rm imp}=0$\mbox{(\textbf{\textcolor[rgb]{0.93,0.69,0.13}{\ \  \raisebox{-0.25ex}{\scalebox{1.3}{$\mathclap{\mathclap{+}\mathclap{\times}}$}}\ \  })}}, $0.3$ eV\mbox{(\textbf{\textcolor[rgb]{0.47,0.67,0.19}{\ \ \raisebox{-0.5ex}{\scalebox{1.75}{$\mathclap{\circ}$}}\ \ })}}, $0.6$ eV\mbox{(\textbf{\textcolor[rgb]{1,0,0}{\ \ \raisebox{-0.2ex}{\scalebox{1.2}{$\mathclap{\times}$}}\ \ })}} and $0.9$ eV\mbox{(\textbf{\textcolor[rgb]{0,0,1}{\ \ \raisebox{-0.2ex}{\scalebox{1.2}{$\mathclap{\square}$}}\ \ })}}. The solid lines show the fitting curves to $y=y_{\infty}(1-c\exp(-x/\lambda_t))$. (b); The values of the fitting parameters for the characteristic length and the saturated values versus the impurity strength. (c) and (d); Same as (a) and (b) for a case of shifted chemical potential, $\mu=E_{\rm F}-1~{\rm eV}$. (e) and (f) The results for $M^y_{\rm ACA}$ and the fitting parameters in the case of shifted chemical potential, $\mu=E_{\rm F}-1~{\rm eV}$. The error bars in (b), (d), and (f) denote the $95~\%$ confidence interval obtained from the standard deviation of the least squares curve fit.}
	\label{fig:fig8}
\end{figure}

    \subsubsection{\label{sec:sec3c1}Thickness dependence}

To further compare the current-induced circular dichroic absorption to the current-induced ACA orbital moment accumulation, we next vary the film thickness $L$ and compute the response (or accumulation) summed over half of the film. This approach mirrors experimental procedures where optical response measurements are carried out as a function of film thickness. To investigate the influence of disorder and go beyond the constant relaxation time approximation, we introduce a random onsite potential along the thickness direction $z$, maintaining translational invariance along the $x$ and $y$ directions. { This introduces a finite decoherence length and localization of the electronic wave function along the $z$-axis.} The potential ensemble used in the numerical calculations is a normal distribution with a width of $V_{\rm imp}$, and we average over more than 20 disorder realizations.

Figure~\ref{fig:fig8}(a) gives the thickness-dependent behavior of current-induced circular dichroism $\alpha^{{\rm surf},yx}_{\rm SR}$ for various strengths of disorder potential. The solid lines represent fits to the data using an exponential expression, \mbox{$y=y_{\infty}(1-c\exp(-x/\lambda_t))$}.
Fig.~\ref{fig:fig8}(b) displays the calculated values of the fitting parameters, including the decay length, $\lambda_t$ as the left vertical axis and the saturated values of the current induced circular dichroism $y_{\infty}$, versus disorder strength $V_{\rm imp}$. We observe a relatively monotonic decrease of the characteristic length with the increasing disorder strength, which can be attributed to the disorder-induced decoherence effect. Similarly, the results demonstrate an inverse proportionality of the saturated value of $\alpha^{{\rm surf},yx}_{\rm SR}$ versus impurity strength. Given its extrinsic nature and proportionality to the relaxation time, this decrease with increasing impurity potential is expected. 

Figure~\ref{fig:fig7}(b) shows that the current-induced $M^y_{\rm ACA}$ is nearly zero at the Fermi level. To study the characteristic length dependence on the disorder and make a comparison between current-induced $\vec{M}_{\rm SR}$ and ACA orbital moment accumulation, we consider a rigid shift of the chemical potential to 1~eV below the Fermi level. Fig.~\ref{fig:fig8}(c) and (e) show the corresponding results for ${\alpha}^{{\rm surf}, yx}_{\rm SR}$ and ${\alpha}^{{\rm surf}, yx}_{\rm ACA}$, versus Pt film thickness for various impurity strengths. The results of the fitting parameters in the case of ${\alpha}^{{\rm surf}, yx}_{\rm ACA}$ are presented in Fig.~\ref{fig:fig8}(d). We observe a characteristic length of about 3~nm that, within the uncertainty remains relatively independent of the disorder. The amplitude of the saturated circular dichroic response decreases with the disorder, as expected. On the other hand, in the case of ${\alpha}^{{\rm surf}, yx}_{\rm ACA}$, Fig.~\ref{fig:fig8}(f) demonstrates that both the characteristic length and the saturated values decrease with the disorder strength.
{In the absence of disorder the fitting yields a characteristic length, $\lambda_t=3.7\pm0.7~{\rm nm}$, where the $\pm$ denotes the $95~\%$ confidence interval obtained from the standard deviation of the least squares curve fit.  This value is close to the mean free path of electrons in bulk Pt, $\ell_{\rm mfp}\approx3.5~ {\rm nm}$, calculated using estimated Fermi velocity of bulk Pt, $v_F=5\times 10^5 ~{\rm m/s}$, \cite{Dutta2017} and broadening value, $\eta=50$ meV.  

For both $\vec{M}_{\rm SR}$ and ACA orbital moment accumulation, the length scale of the thickness dependence of the total surface response is different than the length scale of the spatial profile.  The spatial profile of both $\vec{M}_{\rm SR}$ and ACA orbital moment accumulation (given in Figs.~\ref{fig:fig5}(b) and \ref{fig:fig7}(b,c), respectively) decays to $1/e$ of its maximum value within 5 monolayers (1~nm). In contrast, the length scale governing the thickness dependence ranges from 2~nm to 10~nm, depending on Fermi level and disorder magnitude (see Fig.~\ref{fig:fig8}). This indicates that thickness-dependent measurements of the optical response do \textit{not} provide direct information on the length scale of spatial accumulation. The length scale of the spatial profile is determined by diffusion lengths in the case of bulk diffusive transport, or by lattice spacing in the case of purely surface effects, such as the Edelstein effect. In contrast, the length scale of the thickness-dependent response is governed by the coherence length or mean free path. This is to be expected: the character of an extended state at any position depends on the environment within a coherence length \cite{goedecker1998decay}. Therefore, even a purely surface property will change with thickness up to a mean free path. Measurements of thickness-dependent optical response should, therefore, be interpreted as indicative of the maximum of the diffusion length and mean free path.

In addition to the current-induced surface circular dichroism, the magnetic field, due to the bias current, is also expected to result in a current-induced magnetic moment in the film that can be detected via MOKE or circular dichroic response measurements. Using experimental values of the conductivity, $\sigma=9.4~ {\rm (\mu\Omega \cdot m)^{-1}}$~\cite{Povzner2010}, inter-layer distance, $a=0.2$ nm, and the total (both spin and orbital) magnetic susceptibility, $\chi_m=193\times10^{-12}~ {\rm m^3/mol}$~\cite{Haynes2014} for bulk Pt, the total accumulated magnetic moment induced by the magnetic field in half of a Pt film with $N^{\rm tot}_{{\rm Pt}}$ total number of layers is estimated to be, $\delta M^y_{\rm Ors}/e\mathcal{E}_x^{\rm bias}\approx 4\times10^{-6}(N^{\rm tot}_{\rm Pt})^2\mu_{B}\frac{\rm nm}{\rm eV}$, which becomes comparable to the surface $M^y_{\rm SR}$ when the film thickness exceeds $10^3$ MLs or $200$ nm, which is beyond the range of film thicknesses that is the focus of this work. It should also be noted that optical detection of the induced magnetic moment is limited to the skin depth near the surface. In the case of Pt within the optical spectrum, the skin depth is around $30$ nm, which reduces the total magnetic field-induced magnetization actively interacting with the incident light, further diminishing the contribution of the Oersted effect on the current-induced circular dichroism.
  
\section{\label{sec:sec4}Conclusions}	
In summary, we used {\it ab initio} method to calculate the current-induced circular dichroic absorption in Pt film. We investigated the effects of the chemical potential, thickness, and random onsite disorder strength. We compared the frequency-integrated circular dichroic absorption, $\vec{M}_{\rm SR}$, to the current-induced ACA orbital moment accumulation at the surface of the Pt slab. The results do not show a strong correlation between the two quantities either in amplitude or behavior versus sample variations, such as film thickness and a rigid chemical potential shift. This contrasts with the orbital magnetization in the ground state of a ferromagnet, where $\vec{M}_{\rm SR}$ and the ACA orbital moment are reasonable approximations to the total orbital moment. With a rigid shift in chemical potential, we observed that on average, the current-induced ACA surface orbital magnetization is about one order of magnitude smaller than the surface $\vec{M}_{\rm SR}$. We also find that  $\vec{M}_{\rm SR}$ contains substantial contributions from finite size effects which we estimate to be important for films up to several tens of nanometers. Finally, we show that the length scale governing the thickness dependence of both current induced $\vec{M}_{\rm SR}$ and ACA orbital moment accumulation is set by the mean free path, and is distinct from the length scale describing their spatial profile.

\section{Acknowledgement}
We thank Fei Xue for enlightening discussions and Zachary Levine and Garnett Bryant for carefully reading the manuscript and providing insightful comments. FM acknowledges support under the Cooperative Research Agreement between the University of Maryland and the National Institute of Standards and Technology Physical Measurement Laboratory, Award 70NANB23H024, through the University of Maryland.

\appendix

\section{Computational Methodology}\label{sec.AppA}
The Hamiltonian $\hat{H}_{\vec{R}}$ and overlap $\hat{\mathcal{O}}_{\vec{R}}$, matrix elements of the (001) Pt slab consisting of various thicknesses of Pt  are determined from density functional theory calculations employing the 
OpenMX {\it ab initio} package \cite{OzakiPRB2003,OzakiPRB2004,OzakiPRB2005}.
We adopted Troullier-Martins type norm-conserving pseudopotentials \cite{TroullierPRB1991} with partial core correction. We used a $14\times 14\times 1$ $k$-point mesh for the first Brillouin zone (BZ) integration and an energy cutoff of 500 Ry (1~Ry$\approx$13.6~eV) for numerical integrations in the real space grid. The localized orbitals were generated with radial cutoffs of 7.0~$a_0$ (1~$a_0\approx$~0.529~nm) for Pt \cite{OzakiPRB2003, OzakiPRB2004}.
We used the local spin density approximation (LSDA) \cite{CeperleyPRL1980} exchange-correlation functional as 
parameterized by Perdew and Zunger \cite{PerdewPRB1981}. 
In calculating orbital magnetization for bulk Fe, presented in Fig.~\ref{fig:fig2}(a), we used quadruple-zeta basis sets to get relative convergence with respect to the number of atomic orbitals within the linear combination of atomic orbitals (LCAO) methodology. 
  
 		The velocity operator in Fourier space was calculated using \cite{Lee2018}
		\begin{flalign}			
              \hbar\vec{\hat{v}}_{\vec{k}}&=\frac{\partial \hat{H}_{\vec{k}}}{\partial\vec{k}}-\hat{H}_{\vec{k}}\hat{\mathcal{O}}_{\vec{k}}^{-1}\frac{\partial \hat{\mathcal{O}}_{\vec{k}}}{\partial\vec{k}}-i\left(\vec{\hat{r}}_{\vec{k}}\hat{\mathcal{O}}^{-1}_{\vec{k}}\hat{H}_{\vec{k}}-\hat{H}_{\vec{k}}\hat{\mathcal{O}}^{-1}_{\vec{k}}\vec{\hat{r}}_{\vec{k}}\right), \label{Eq.A1}  
		\end{flalign}
  where the $\vec{k}$-derivatives are obtained using
		\begin{flalign}			
              \frac{\partial{\hat{C}}_{\vec{k}}}{\partial\vec{k}}&=
              i\sum_{\vec{R}}\vec{R}\hat{C}_{\vec{R}}e^{i\vec{k}\cdot\vec{R}}.
		\end{flalign}
		where $\hat{C}$ can represent either the Hamiltonian or the overlap matrix. The position operator within the unit cell is calculated using $\vec{\hat{r}}_{\vec{k}}=\sum_{\vec{R}}\vec{\hat{r}}_{\vec{R}}e^{i\vec{k}\cdot\vec{R}}$, where, \mbox{$\vec{{r}}_{\vec{R}}^{\ I\mu,J\nu}=\vec{r}_I{\mathcal{O}}^{\ I\mu,J\nu}_{\vec{R}}+\vec{{r}}^{\ I\mu,J\nu}_{a,\vec{R}}$}. Here $\mu,\nu$ represent the atomic orbital basis sets, and $I,J$ are the atomic indices inside the unit cell. The position operator within the atoms is evaluated using
		\begin{flalign}
			\vec{{r}}^{\ I\mu,J\nu}_{a,\vec{R}}=\int d\vec{r}~\phi_{I\mu}(\vec{r})\vec{r}\phi_{J\nu}(\vec{r}-\vec{r}_{J}-\vec{R}+\vec{r}_{I}),
		\end{flalign}
		with, $\phi_{I\mu}(\vec{r})$, the atomic orbital basis functions for atom $I$.

\section{Finite Size Effects}\label{sec.AppB}

In this section, we evaluate the thickness-dependent dichroic absorption for a single-band model. In this model, {\it all} transitions between states are within the same bulk band, and it illustrates the inter-subband contribution to the dichroic absorption. With this single-band model, we can demonstrate the bound given by Eq.~\eqref{Eq.omegaL} of the main text and the relative contribution of different terms entering into the dichroic absorption.

The model consists of a thin film of thickness $L$ along the $z$-direction; $L=Na$ where $N$ is the number of layers and $a$ is the interlayer spacing. The closed boundary condition leads to standing waves along the $z$-direction with a mode index labeled by $n$. Assuming that the wave functions are separable, we have
\begin{flalign}			
\psi_{n}({\vec r})=\sqrt{\frac{2}{L}}\sin(\frac{n\pi}{L}z)~\phi_{n}(x,y).
\end{flalign}
In what follows, we refer to the $\sin(\frac{n\pi}{L}z)$ part as the envelope wave function.

As shown in Eq.~\eqref{eq:alpha_eq} of the main text, the term in the response function that enters into the dichroic absorption is ${\rm Im}(\chi^{{\rm surf}}_{xz;x})$.  Our aim is to evaluate this term for this model.  We re-write ${\rm Im}(\chi^{{\rm surf}}_{xz;x})$ using the relation between velocity and position: $\hbar{{v}}^z_{mn}=i(\varepsilon_m-\varepsilon_n){{z}}_{mn}$, obtaining  
\begin{eqnarray}
{\rm Im}(\chi^{{\rm surf}}_{xz;x})&=&{\frac{e^2}{4V\omega}}\sum_{mn}{\rm Re}\left({{{v}}^{x,{\rm full}}_{mn}{{z}}^{{\rm half}}_{nm}}+{{{v}}^{x,{\rm half}}_{mn}{{z}^{\rm full}}_{nm}}\right)\nonumber \\&&~~~~~~
\delta({\varepsilon_{n}-\varepsilon_{m}-\hbar\omega})(\delta f^{x}_{m}-\delta f^{x}_{n})\label{Eq.C2}.
\end{eqnarray}
The superscripts ``full'' and ``half'' refer to the matrix element region of integration: indicating whether the spatial integral along the film-normal direction extends over the full or half-film.

The matrix elements in Eq.~\eqref{Eq.C2} can be evaluated explicitly. Their important feature is the selection rules which constrain the values of $n$ and $m$:
\begin{eqnarray}
{{v}}^{x,{\rm full}}_{mn}& \propto \delta_{n,m}\label{Eq.C3},\\
{{z}}^{{\rm full}}_{m,n}&\propto \delta_{n-m,\ell}\label{Eq.C4},
\end{eqnarray}
where $\ell=2p+1$ is an odd integer. Eq.~\eqref{Eq.C3} reflects the orthogonality of the envelope wave functions, while Eq.~\eqref{Eq.C4} is a statement of the optical selection rule.  Eq.~\eqref{Eq.C3} immediately implies that the contribution of the first term in Eq.~\eqref{Eq.C2} is zero, as transitions only occur between states $n\neq m$. Note that in a multi-band model, the first term {\it can} contribute to the circular dichroic absorption due to bulk interband transitions. However, the inter-subband contribution to dichroic absorption is entirely due to the second term of Eq.~\eqref{Eq.C2} ({\it i.e.,} the ${v}_{mn}^{x,{\rm half}} {z}_{mn}^{\rm full}$ term). In the full model, evaluating the difference in the two terms of Eq.~\eqref{Eq.C2} can therefore provide an indication of the relative contribution of inter-subband transitions to the total circular dichroic absorption. 

We next evaluate the $\delta$-function constraint of Eq.~\eqref{Eq.C2}. The energy levels of the finite film can be estimated by taking the band structure of the periodic system and constructing a supercell of size $N$, which reduces the Brillouin zone by $1/N$. This effectively ``folds'' the bands $N$ times and the resulting energy values at $k_z=0$ provide an estimate of the energy levels of the finite film. The energy spacing between nearby modes $n$ and $m$ of the film can then be approximated by
\begin{eqnarray}
    \varepsilon_{n}-\varepsilon_{m} &\approx& \frac{\pi}{Na} \left(n-m\right)\left.\frac{\partial E}{\partial k_z} \right|_{k_z=k_n}  \\ 
    &=& \frac{\hbar \pi}{L} \left(n-m\right) v^z_n,\label{eq:de_est}
\end{eqnarray}
where, $E(k_z)$ is the band structure for the periodic system, $k_n = n\pi/L$, and $v^z_n=(1/ \hbar) ~ \partial E/\partial k_z|_{k_z=k_n}$ is the group velocity. Combining the requirement that $\varepsilon_{n}-\varepsilon_{m}=\hbar\omega$, the selection rule from Eq.~\eqref{Eq.C3} and the energy difference from Eq.~\eqref{eq:de_est} yields
\begin{eqnarray}
n-m=\ell\approx \frac{\omega L}{\pi v^z_n}. \label{Eq.C5}
\end{eqnarray}
In general, inter-subband transitions are most prominent for small values of the integer $\ell$.  The velocity that ultimately enters in Eq.~\eqref{Eq.C5} is the Fermi velocity, as we are focused on current-induced changes to the absorption. Eq.~\eqref{Eq.C5} demonstrates several important properties of circular dichroic absorption: ({i}) The energy of inter-subband transitions is inversely proportional to layer thickness. The magnitude of inter-subband absorption, therefore, increases linearly with thickness due to the $1/\omega$ factor in Eq.~\eqref{Eq.C2}. ({ii}) The frequency and thickness dependence varies with the product $\omega L$ (see Figs.~6(b) and 6(d) of the main text). ({iii}) The group velocity at the Fermi level, $v^z_F$, determines the relevant scale of $\omega L$ for which inter-subband transitions are important, in agreement with the heuristic argument given in the main text. These features provide motivation for the form of the dichroic absorption given in Eq.~\eqref{Eq.dich_opt_decomp} of the main text. 

A physical picture of the inter-subband circular dichroic absorption can be obtained by analyzing the remaining matrix element ${{v}}_{nm}^{x,{\rm half}}$ in more detail. Integrating this matrix element over the upper or lower half of the film yields a positive or negative value of $v_x$, respectively. The relative sign comes from the different sign of envelope functions on the upper and lower halves of the slab, which in turn follows from the opposite parity of $n$ and $m$. Hence, for a spatially uniform electric field oscillating in the $z$-direction, an equal and opposite flow along the $x$-direction is induced in the upper and lower half of the slab. This current flow is $90^{\circ}$ out-of-phase with the applied field, hence absorbs the energy of circularly polarized light. The spatial extent of the induced current is set by the envelope function. The direction of the current flow is odd in $k_x$; however, applying an electric field along $x$ modifies the distribution function and leads to a net nonzero value over one-half of the film.

\section{Fitting Method and Results}\label{sec.AppC}
The method used to calculate the uncertainty in the fitting process for Fig.~\ref{fig:fig8} involves two main steps. First, the model parameters of the fitting function, $F(L)=F({\infty})(1-c\exp(-L/\lambda_t))$, are estimated using a nonlinear least-squares approach. $F$ represents the quantities $\alpha_{\rm SR}^{yx}$ and $\alpha_{\rm SR}^{yx}$. Then, the uncertainties for the estimated parameters, $F({\infty})$, $c$, and $\lambda_t$ are calculated using 95\% confidence intervals using the standard deviation from the least squares curve fitting. This approach allows for quantifying the uncertainty associated with the estimated model parameters. For $c\approx1$, the model may be interpreted as describing surface-driven diffusion. 

The fitting parameters for the results presented in Figs.~\ref{fig:fig8} are listed in Tables, \ref{table:1}, \ref{table:2} and \ref{table:3}. 
  \begin{table}[h!]
\centering
\begin{tabular}{ c c c c } 
 \hline\hline
 $V_{\rm imp}$ [eV] & $\alpha_{\rm SR}^{yx}({\infty})/\mu_{B}[{\rm nm/eV}]$ & $c$  & $\lambda_t$ [nm]\\ [0.5ex] 
 \hline
 0 & 9.16$\pm$1.88 & 0.97$\pm$0.065 & 10.1$\pm$4.55 \\ 
 0.3 & 4$\pm$0.35 & 1.01$\pm$0.18 & 4.7$\pm$1.7 \\
 0.6 & 2.04$\pm$0.9 & 0.96$\pm$0.18 & 3.4$\pm$0.9 \\
 0.9 & 1.5$\pm$0.14 & $-1.34$$\pm$1.55 & 1.9$\pm$1.65 \\ 
 \hline \hline
\end{tabular}
\caption{Fitting parameters for current induced circular dichroism versus Pt film thickness at $\mu=E_{\rm F}$. Here $\alpha_{\rm SR}^{yx}({\infty})$ corresponds to the saturated self-rotating component of the orbital moment with respect to the Pt thickness.}
\label{table:1}
\end{table}

  \begin{table}[h!]
\centering
\begin{tabular}{c c c c} 
 \hline\hline
 $V_{\rm imp}$ [eV] & $\alpha_{\rm SR}^{yx}({\infty})/\mu_{B}[{\rm nm/eV}]$ & $c$  & $\lambda_t$ [nm]\\ [0.5ex] 
 \hline
 0 & $-4.4$$\pm$2.1 & 3.67$\pm$3.93 & 2.6$\pm$2.35 \\ 
 0.3 & $-0.92$$\pm$1.2 & 4.43$\pm$6.45 & 5.4$\pm$6.05 \\ 
 0.6 & 0.56$\pm$0.18 & $-3.15$$\pm$1.56 & 2.9$\pm$2.05 \\
 0.9 & 0.78$\pm$0.16 & $-0.91$$\pm$1.1 & 2.8$\pm$4.4 \\ 
 \hline \hline
\end{tabular}
\caption{Fitting parameters for current induced circular dichroism versus Pt film thickness at $\mu=E_{\rm F}-1$ eV. }
\label{table:2}
\end{table}

  \begin{table}[h!]
\centering
\begin{tabular}{c c c c} 
 \hline\hline
 $V_{\rm imp}$ [eV] & $\alpha_{\rm ACA}^{yx}({\infty})/\mu_{B}[{\rm nm/eV}]$ & $c$  & $\lambda_t$ [nm]\\ [0.5ex] 
 \hline
 0 & 0.61$\pm$0.03 & 1.4$\pm$0.2 & 3.7$\pm$0.7 \\ 
 0.3 & 0.33$\pm$0.025 & 1.8$\pm$0.6 & 2.6$\pm$0.85 \\ 
 0.6 & 0.08$\pm$0.03 & 4.5$\pm$6.65 & 1.6$\pm$1.55 \\
 0.9 & $-0.05$$\pm$0.01 & $-3.4$$\pm$7.75 & 1.4$\pm$2.0 \\ 
 \hline \hline
\end{tabular}
\caption{Fitting parameters for current induced circular dichroism versus Pt film thickness at $\mu=E_{\rm F}-1$ eV. Here $\alpha_{\rm ACA}^{yx}({\infty})$ corresponds to the saturated self-rotating component of the orbital moment with respect to the Pt thickness.}
\label{table:3}
\end{table}

\section{Connection to the circular photogalvanic effect}\label{sec.AppD}
In this section, we provide the analytical derivation of the pumped direct current due to the circular photogalvanic effect and compare it to the expression for current-induced circular dichroism. In response to a position-dependent electric field, the direct current that is given by \cite{Sipe2000} 
\begin{flalign}
{\rm I}^{\gamma}_{\rm DC}&=\sum_{IJ\alpha\beta}{\rm Re}({\rm \mathcal{E}}_{I\alpha}{\rm \mathcal{E}}^{\mbox{*}}_{J\beta}\xi_{\alpha\beta,\gamma}^{IJ}),\\
&=\sum_{IJ\alpha\beta}{\rm Re}({\rm \mathcal{E}}_{I\alpha}{\rm \mathcal{E}}^{\mbox{*}}_{J\beta}){\rm Re}(\xi_{\alpha\beta,\gamma}^{IJ})-{\rm Im}({\rm \mathcal{E}}_{I\alpha}{\rm \mathcal{E}}^{\mbox{*}}_{J\beta}){\rm Im}(\xi_{\alpha\beta,\gamma}^{IJ}),\nonumber
\end{flalign}
where the first and second terms describe the shift and the circular-photogalvanic-effect currents, respectively, and
\begin{flalign}
\xi_{\alpha\beta,\gamma}^{IJ}&=\sum_{mn}\frac{e\pi v^{\gamma}_{nn}}{i\eta\hbar^2\omega^2}\left[v^{I\alpha}_{nm}v^{J\beta}_{mn}\delta(\varepsilon_{n}-\varepsilon_{m}+\hbar\omega)\right.\nonumber\\
&+\left.v^{I\alpha}_{mn}v^{J\beta}_{nm}\delta(\varepsilon_{n}-\varepsilon_{m}-\hbar\omega)\right](f_m-f_n).
\end{flalign}
The circular-photogalvanic-effect component of the nonlinear response can be rewritten as
\begin{flalign}
{\rm Im}&(\xi_{\alpha\beta,\gamma}^{IJ})=\frac{\pi e^3}{\eta\hbar\omega}\sum_{mn}{\rm Im}(v^{I\alpha}_{nm}v^{J\beta}_{mn})\delta(\varepsilon_{n}-\varepsilon_{m}+\hbar\omega)\\
&\times\left(v_{nn}^{\gamma}\frac{f(\varepsilon_n+\hbar\omega)-f(\varepsilon_n)}{\hbar\omega}-v_{mm}^{\gamma}\frac{f(\varepsilon_m)-f(\varepsilon_m-\hbar\omega)}{\hbar\omega}\right).\nonumber
\end{flalign}
In the limit of $\hbar\omega\ll kT$, we obtain
\begin{flalign}
{\rm Im}(\xi^{IJ}_{\alpha\beta,\gamma})&=\frac{\pi e^3}{2\hbar\omega\eta}\sum_{nm}{\rm Im}(v^{I\alpha}_{nm}v^{J\beta}_{mn})\delta(\varepsilon_{n}-\varepsilon_{m}+\hbar\omega)\nonumber\\&[v^{\gamma}_{nn}f'(\varepsilon_{n})-v^{\gamma}_{mm}f'(\varepsilon_{m})],
\end{flalign} 
which is identical to the imaginary part of the circular dichroic response
\begin{flalign}
{\rm Im}(\chi^{IJ}_{\alpha\beta;\gamma})&=\frac{\pi e^2}{2\hbar\omega\eta}\sum_{mn}{\rm Im}({{v}}^{ I\alpha}_{mn}{{v}}^{J\beta}_{nm})\delta(\varepsilon_{n}-\varepsilon_{m}+\hbar\omega)\nonumber\\
&(v^{\gamma}_{nn}f'(\varepsilon_{n})-v^{\gamma}_{mm}f'(\varepsilon_{m})).
\end{flalign}

		\bibliographystyle{apsrev4-1}
		\bibliography{ref.bib}
		
\end{document}